\documentclass[prb,twocolumn,showpacs,superscriptaddress,floatfix]{revtex4-1}
\usepackage{graphicx}
\usepackage{amssymb,amsmath}
\usepackage[colorlinks,citecolor=blue,urlcolor=blue]{hyperref}

\usepackage{color}

\begin{document}

\title{
Spin and charge dynamics of a quasi-one-dimensional antiferromagnetic metal}

\author{Marcin Raczkowski}
\affiliation{Department of Physics and Arnold Sommerfeld Center for Theoretical Physics, 
             Ludwig-Maximilians-Universit\"at M\"unchen, D-80333 M\"unchen, Germany}
\author{Fakher F. Assaad}
\affiliation{Institut f\"ur Theoretische Physik und Astrophysik,
             Universit\"at W\"urzburg, Am Hubland, D-97074 W\"urzburg, Germany}
\author{Lode Pollet}
\affiliation{Department of Physics and Arnold Sommerfeld Center for Theoretical Physics,
             Ludwig-Maximilians-Universit\"at M\"unchen, D-80333 M\"unchen, Germany}

\date{\today}

\begin{abstract}
We use quantum Monte Carlo simulations to study a finite-temperature dimensional-crossover-driven 
evolution of spin and charge dynamics in an anisotropic two-dimensional system of
weakly coupled Hubbard chains with a half-filled band. 
The low-temperature behavior of the charge gap indicates a crossover between two distinct energy scales:
a high-energy  one-dimensional (1D) Mott gap due to the umklapp process 
and a low-energy gap which stems from long-range antiferromagnetic (AF) spin fluctuations.
Away from the 1D regime and at temperature scales above the charge gap, the emergence of a zero-frequency 
Drude-like feature in the interchain optical conductivity $\sigma_{\perp}(\omega)$ 
implies the onset of a higher-dimensional metal. 
In this metallic phase, enhanced quasiparticle scattering off finite-range AF spin  
fluctuations results in incoherent single-particle dynamics.
The coupling between spin and charge fluctuations is also seen in the spin dynamical structure 
factor $S({\pmb q},\omega)$ displaying damped spin excitations (paramagnons) close to the 
AF wave-vector ${\pmb q}=(\pi,\pi)$ and particle-hole continua near 1D momentum transfers 
spanning quasiparticles at the Fermi surface. 
We relate our results to the charge deconfinement in quasi-1D organic Bechgaard-Fabre salts.
\end{abstract}

\maketitle

\section{Introduction}

Dimensional crossovers are interesting because there are relevant couplings between the lower dimensional objects 
which dramatically alter the physical properties of the system.~\cite{Kivelson00} 
The renormalization at low energy and long wavelengths is therefore higher-dimensional, 
but the question of how remnants of the one-dimensional (1D) physics remain visible when the couplings get stronger 
is largely left open.
One would expect the 1D characteristics to survive when probing the system at high energy or when the temperature 
is larger than the interchain coupling but lower than the typical scales of the 1D problem. 
Below we give three examples of experiments in different fields whose complete description requires 
a deeper understanding of dimensional crossover phenomena.

Cold atoms in optical lattices provide a controllable and flexible model realization of correlated quantum 
systems and offer a clean setup  to probe their dynamics.~\cite{Bloch08,Bloch12}
For example, a variable strength of the interchain coupling in a recent realization of a tunable optical 
lattice comprising weakly coupled 1D chains,~\cite{Greif13}  allows one to study the impact of a dimensional 
crossover on antiferromagnetic (AF) spin correlations and stimulated a renewed interest in low-dimensional 
quantum many-body physics.~\cite{Sciolla13,Heikkinen13,Heikkinen14,Imriska14,Lode14}

Other systems to explore the interplay between low-dimensional quantum dynamics 
and electron correlations are quasi-1D organic Bechgaard-Fabre salts.~\cite{Giamarchi04} 
A rich variety of phenomena in their global temperature-pressure phase diagram has been ascribed to 
a decreasing degree of dimerization, i.e., the strength of the umklapp process,~\cite{Giamarchi91} 
and increasing electronic dimensionality with applied pressure which triggers 
a metal-insulator transition.~\cite{Vescoli98,Pashkin10}
Whether the nature of this higher-dimensional metallic phase and its low-energy excitations can be accounted 
for by a Fermi liquid (FL) theory is the key question in the physics of these compounds.

Finally, fingerprints of a strongly anisotropic metallic phase 
have been observed in the proximity to the AF ground state in high-$T_c$ cuprate superconductors.~\cite{Vojta09a} 
In this case, a dimensional reduction is driven by an inhomogeneous self-organization 
of doped holes which condense into arrays of parallel stripes separating regions  
with enhanced AF spin correlations. 
Since the charge dynamics occurs mainly along the stripes, 
it can be effectively described by a quasi-1D model 
in which the transport across the stripes is incoherent.~\cite{Fradkin14}

From the theoretical point of view, the complexity of a dimensional crossover in coupled 
1D Hubbard chains comes from single- and two-particle processes generated by the interchain
coupling.~\cite{Gogolin_book}
On the one hand, the crossover in metallic chains is easily induced by the  
interchain \emph{one-particle} hopping process thus replacing the Luttinger liquid (LL) behavior
with a conventional FL metal.~\cite{Castellani94,Boies95,Kopietz95,Arrigoni99} 
On the other hand, the umklapp-induced Mott gap in a half-filled band 
makes the problem more difficult due to the enhanced relevance of \emph{two-particle} fluctuations: 
binding of particle-hole pairs from neighboring chains generates a finite AF superexchange coupling $J_{\perp}$ and
may induce an onset of the broken-symmetry spin-density-wave (SDW)  phase.~\cite{Brazovskii85}
In contrast, when the single-particle tunneling $t_{\perp}$ overcomes the magnetic coupling 
$J_{\perp}$, it drives a metal-insulator transition thus leading to the formation of 
a higher-dimensional Fermi surface (FS).~\cite{Bier01,Ess02,Tsu07, Rib11,Mou11}

Recently, we have studied the effects of interchain coupling between the half-filled Hubbard chains 
in two limiting cases in the effective \emph{zero}-temperature regime.~\cite{Raczkowski12,Raczkowski13}  
First, the dimensional crossover dominated by single-particle tunneling 
can be examined  with a cluster extension of the dynamical mean-field theory 
(CDMFT).~\cite{Georges96,Kotliar01} 
On the one hand, CDMFT neglects most of the spatial magnetic fluctuations 
restricted to the cluster size. On the other hand, it captures well umklapp 
scattering and thus reproduces the density-driven Mott transition in 
the 1D Hubbard model.~\cite{Bolech03,Capone04}
The CDMFT scenario of the dimensional-crossover-driven metallic phase 
with a FS broken into pockets is plausible when the low-temperature SDW instability 
of the nested FS is eliminated, e.g., by a sufficiently large 
geometrical frustration; a similar broken FS has also been 
found in the spinless fermion model.~\cite{Berthod06} 

Second, the two-particle crossover and a dynamical confinement which occurs, for instance, in
a system of coupled spin-1/2 Heisenberg chains,~\cite{Schultz96,Essler97} 
can be addressed with lattice quantum Monte Carlo (QMC) simulations by tracking 
the evolution of elementary spin excitations.~\cite{Raczkowski13} 
Indeed, the interchain AF superexchange coupling confines fractionalized 
excitations (spinons) of the individual 1D chains back to form $S = 1$ magnons, 
the Goldstone modes of the broken continuous SU(2) symmetry group.  
This gives rise to low-frequency spin waves which decay into a two-spinon 
continuum at high energies seen in the QMC dynamical spin structure factor 
$S({\pmb q},\omega)$.~\cite{Raczkowski13} 
Such a dual nature of magnetic excitation spectra has been resolved in the 
inelastic neutron scattering data on weakly coupled $S=1/2$ chains 
of BaCu$_2$Si$_2$O$_7$ and KCuF$_3$.~\cite{Zhe00, Zhe01, Lake05}

The present work is aimed at investigating the regime in which the single- and two-particle 
interchain fluctuations intertwine and contribute to the low-frequency dynamics of the system. 
In a correlated metal, a mutual interaction between finite-range spin fluctuations and 
fermionic quasiparticles (QPs) introduces damping of the collective spin excitations (paramagnons) 
which decay into a continuum of independent electron-hole pairs.~\cite{Hertz76,Moriya_book,Millis93}
Concurrently, dressing of QPs with a cloud of spin fluctuations enhances the QP scattering 
rate and leads to a partial depletion of the single-particle spectral weight 
at the Fermi level.~\cite{Kampf90,Kampf90a} 
This non-Fermi-liquid behavior is frequently reflected in anomalous transport and thermodynamic 
properties of nearly AF metals and its theoretical description has proven to be a 
challenge.~\cite{Abanov03,Moriya03,Kontani08,Das14}

To treat the single-particle tunneling and dynamically generated interchain superexchange interaction 
on equal footing, we use an auxiliary-field QMC algorithm.~\cite{BSS} 
We characterize the low-frequency excitations in the quasi-1D metallic 
phase by looking at both momentum-resolved single-particle spectral function and
 two-particle quantities: frequency- and polarization dependent optical conductivity 
as well as spin and charge dynamical structure factors.
The paper is organized as follows: After introducing the model and method in 
Sec.~\ref{Model}, we present our results in Sec.~\ref{Results}, and 
then conclude in Sec.~\ref{Conclusions}.

\section{Model and QMC method} 
\label{Model}

We consider the Hubbard model on a square lattice with an anisotropic hopping at half-filling,
\begin{equation}
H-\mu N=-\sum_{\pmb{ij},\sigma}t^{}_{\pmb{ij}}
   c^{\dag}_{{\pmb i}\sigma}c^{}_{{\pmb j}\sigma} +
   U\sum_{\pmb i}n^{}_{{\pmb i}\uparrow}n^{}_{{\pmb i}\downarrow} 
   -\mu\sum_{\pmb i,\sigma}n_{{\pmb i}\sigma},
\label{Hubb}
\end{equation}
where the hopping $t_{\pmb{ij}}$ is $t$ ($t_{\perp}$) on the intrachain (interchain)
bonds, extended by a \emph{diagonal} next-nearest neighbor hopping $t'=-t_{\perp}/4$.
Thus, the corresponding single-particle dispersion relation reads,
\begin{equation}    
\varepsilon_{\pmb k}= -2(t\cos k_x +t_{\perp}\cos k_y) + t_{\perp} \cos k_x\cos k_y. 
\end{equation}
The diagonal hopping $t'$ brings about geometrical frustration thus reducing 
FS nesting properties responsible for the onset of the long-range AF phase in coupled 
spin $S=1/2$ Heisenberg chains.~\cite{Affleck94,Affleck96,Sandvik99,Kim00} 
To account for broken particle-hole symmetry and force a half-band
filling away from the 1D limit, we adjust the chemical potential $\mu$.
Previously,~\cite{Raczkowski13} we have shown that the Hubbard model~(\ref{Hubb}) 
with $U/t=3$ develops N\'eel order in the presence of any infinitesimally small 
interchain coupling in the effective zero-temperature limit. 
Here, we consider a smaller value of the Coulomb interaction $U/t=2.3$ and study 
\emph{finite}-temperature properties of the emergent quasi-1D ($t_{\perp}/t\leq 0.3$.) metallic phase.

For this purpose, we use a finite-temperature implementation of the auxiliary-field 
QMC algorithm  (see Ref.~\onlinecite{Assaad08_rev} and references therein) which 
allows one to compute the expectation value of an observable $O$ in the grand-canonical ensemble:
\begin{equation}
\langle O\rangle=\frac{{\mbox{Tr}}\bigl [e^{-\beta(H-\mu N)}O\bigr ]}{{\mbox{Tr}}\bigl [e^{-\beta(H-\mu N)}\bigr ]}.
\end{equation}
It is based on a path integral formulation of the partition function which maps a quantum system 
in $d$ spatial dimensions onto a $d+1$-dimensional classical problem with an additional 
imaginary time dimension $\beta=1/T$. 
The essence of the QMC method is to separate the one-body kinetic $H_t$ and two-body Hubbard interaction $H_U$ terms 
with the help of the Trotter decomposition,  
\begin{equation}
 e^{-\Delta \tau \left( H_U + H_t \right) } \simeq 
e^{-\Delta \tau H_U } e^{ -\Delta  \tau H_t }. 
\end{equation}
We have used a fixed small discretization of the temporal axis $\Delta \tau t = 1/6$. 
This introduces an overall  controlled systematic error of order $(\Delta  \tau)^2$. 
We have opted for a  discrete, Ising,  Hubbard-Stratonovitch  field coupling to the $z$-component 
of the magnetization.  The QMC simulations were performed for lattice sizes ranging 
from $L=8$ to $L=20$ and in the broad range of temperatures $t/5\le T\le t/30$ restricted by 
the QMC minus-sign problem. 
In addition to equal-time  observables, the QMC method provides access to imaginary time displaced 
correlation functions. 
We use a stochastic version of the maximum entropy method~\cite{Beach04a} to analytically continue 
the imaginary-time QMC data and to extract the real-frequency single- and two-particle 
excitation spectra.

\section{Results}
\label{Results}

We proceed now to present our QMC results for the Hubbard model (\ref{Hubb}). We first consider 
static quantities and then study single- and two-particle dynamical correlation functions.

\subsection{Uniform charge susceptibility and static spin structure factor}
\label{Static}

\begin{figure}[b!]
\begin{center}
\includegraphics[width=0.45\textwidth]{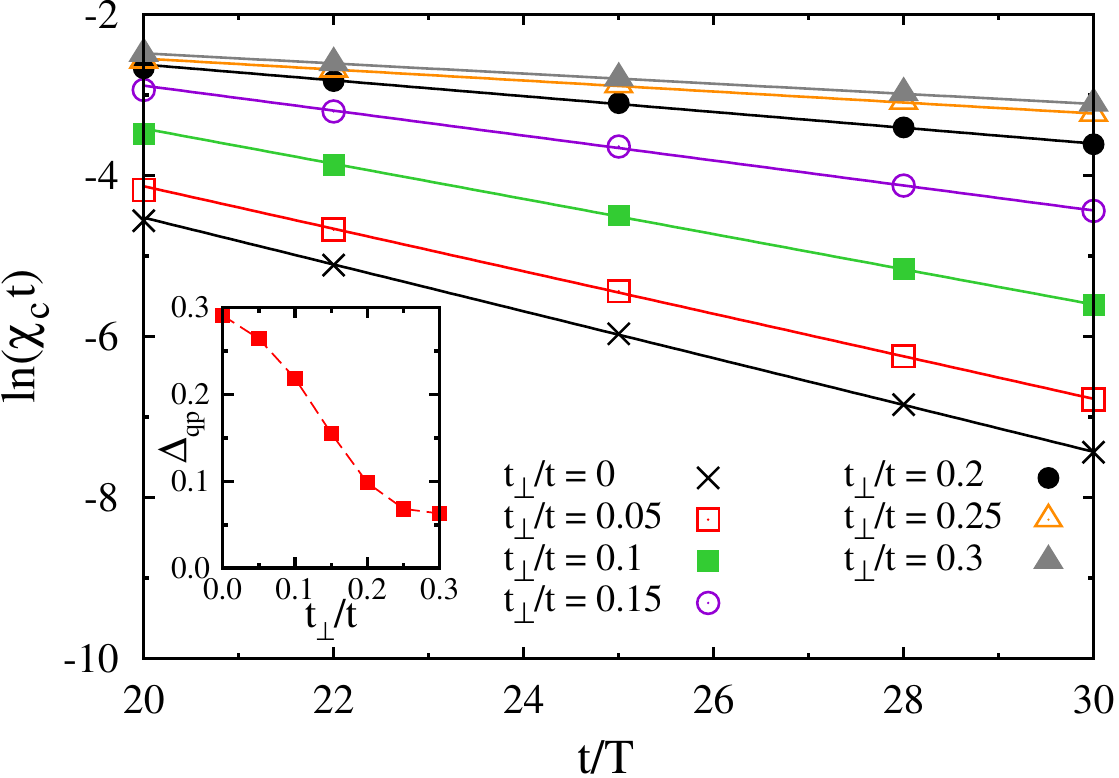}
\end{center}
\caption
{(Color online)
Uniform charge susceptibility $\ln(\chi_c t)$  vs. $1/T$ for several values of the 
interchain hopping $t_{\perp}$ found on a $16\times 16$ lattice. 
The inset shows the evolution of the QP gap $\Delta_{qp}$ on increasing $t_{\perp}$ 
extracted from a linear fit.
}
\label{c00}
\end{figure}

Due to the relevance of umklapp scattering in the 1D limit, the single-particle imaginary time Green's function,
$G({\pmb r}=0,\tau)=\langle c_{{\pmb r}}^{}(\tau) c_{{\pmb 0}}^{\dag} \rangle$,   
follows for large $\tau t\gg 1$ an exponential decay $\propto e^{-\Delta_{qp}\tau}$, where 
$\Delta_{qp}$ is the QP gap. The latter might be extracted from the uniform charge susceptibility,
\begin{equation}
\chi_c=\frac{\beta}{L^2}\Bigl(\langle N^2\rangle- \langle N\rangle^2\Bigr),
\end{equation}
where $N$ is the particle number operator. 
At low temperatures one expects $\chi_c\propto e^{-\Delta_{qp}/T}$. 
In order to quantify the behavior of the QP gap upon increasing $t_{\perp}$, 
we plot in Fig.~\ref{c00} $\ln(\chi_c t)$  vs. $1/T$ for several values of the interchain hopping. 
In this way, the activated behavior is reflected in the slope of the curves given by $\Delta_{qp}$: 
a decreased negative slope signals the reduction of the magnitude of $\Delta_{qp}$. 
A rough estimate of the QP gap $\Delta_{qp}$ extracted from a linear fit of the data points 
indicates that it does not close but shows an inflection point around $t_{\perp}/t=0.15$, 
see the inset in Fig.~\ref{c00}.

\begin{figure}[t!]
\begin{center}
\includegraphics[width=0.45\textwidth]{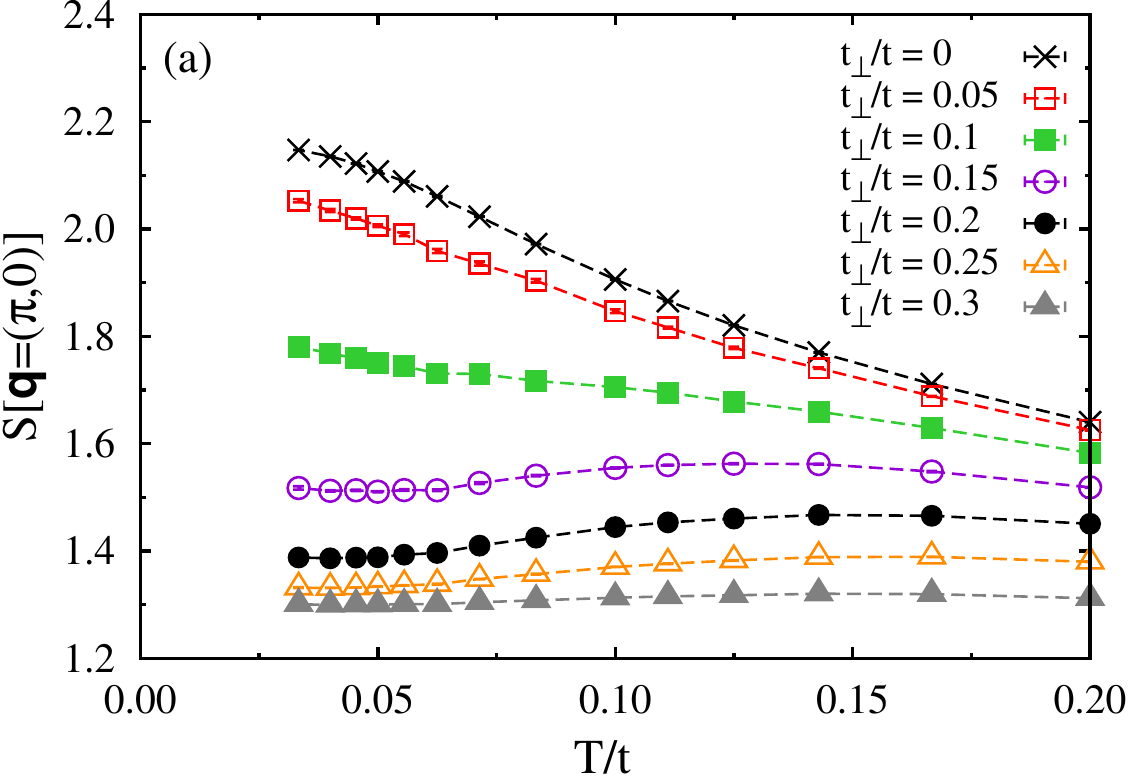}\\
\includegraphics[width=0.45\textwidth]{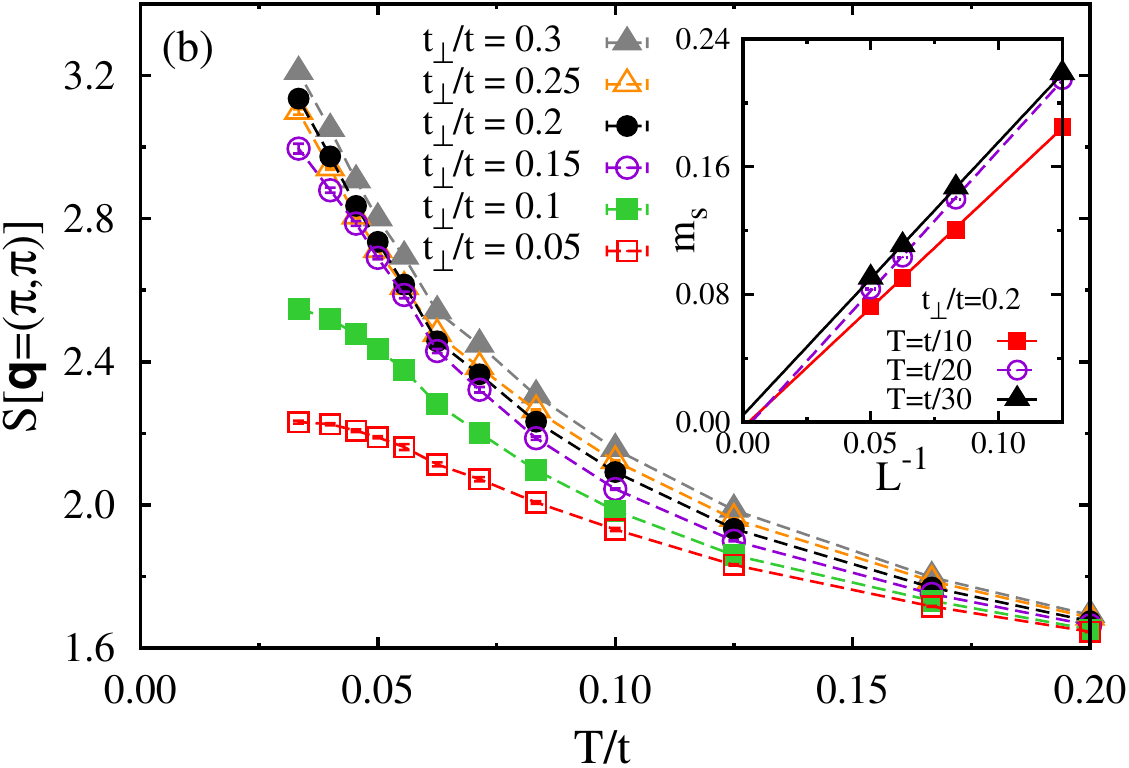}
\end{center}
\caption
{(Color online)
Temperature dependence of the static spin structure factor $S({\pmb q})$
found on a 16$\times 16$ lattice:
(a) ${\pmb q}=(\pi,0)$ and
(b) ${\pmb q}=(\pi,\pi)$.
The inset shows the finite-size scaling of the staggered magnetic moment $m_s$ for 
$t_{\perp}/t=0.2$ at $T=t/10$, $t/20$, and $t/30$.
}
\label{S_T}
\end{figure}

The existence of the inflection point is suggestive of a crossover in the origin of the charge gap.
Numerical evidence for this conjecture is provided in Fig.~\ref{S_T} showing the 
Fourier transform of equal-time spin-spin correlations,
\begin{equation}
S({\pmb q})=\frac{4}{3}
    \sum_{{\pmb r}}e^{i{\pmb q}\cdot{\pmb r}}\langle {\pmb S}({\pmb r})\cdot{\pmb S}({\pmb 0}) \rangle.
\end{equation}
On the one hand, the low-$T$ enhancement of $S(\pi,0)$  for $t_{\perp}/t\leq 0.1$ in Fig.~\ref{S_T}(a)  
indicates the predominance of spin-spin correlations along the chains thus confirming 
the relevance of umklapp scattering. 
On the other hand, the low-$T$ increase of the staggered spin structure factor $S(\pi,\pi)$ 
shown in Fig.~\ref{S_T}(b) suggests the formation of significant AF spin correlations.  
To further study the spin-spin correlations,   
we plot in the inset of Fig.~\ref{S_T}(b) the finite-size scaling of the staggered magnetic moment,
\begin{equation}
m_s=\lim_{L\to\infty}\sqrt{\frac{S(\pi,\pi)}{L^2}},
\end{equation}
at $t_{\perp}/t=0.2$. As is apparent, $m_s$ is consistent with a finite value \emph{below} our lowest temperature 
$T=t/30$, thus marking the effective zero-temperature regime in our finite-size systems.~\footnote{
At sufficiently low temperatures, a finite system develops local moments 
when the QMC lattice size becomes smaller than the AF spin correlation length.}

Hence, on the basis of the static quantities, one can conclude in the $T=0$ limit 
the onset of a higher-dimensional insulating phase gapped out by long-range AF spin fluctuations. 
However, a strong reduction of the QP gap $\Delta_{qp}$ upon increasing $t_{\perp}$ implies 
that for this value of the Coulomb repulsion $U/t=2.3$, the onset of the insulating state occurs 
at lower temperatures in comparison to the 1D Mott gap. 
This opens up a possibility to study finite, but still low-enough to avoid dominant thermal broadening 
effects, temperature properties of the quasi-1D metallic state in the dimensional crossover.

\subsection{Optical conductivity}

A response function particularly suitable for investigating transport 
properties in the anisotropic system such as weakly coupled 1D Hubbard chains is the frequency- 
and polarization-dependent optical conductivity $\sigma_{\alpha}(\omega)$:
it allows one to resolve a distinct behavior of the charge dynamics 
along $\sigma_{\parallel}(\omega)$ and perpendicular $\sigma_{\perp}(\omega)$ to 
the chains.
Noting that, experimentally, the dimensionality is controlled not only by the physical or
chemical pressure, which changes the ratio of the interchain
transfer integral to the intrachain one, but also by the energy
scale used in the measurement, the knowledge of $\sigma_{\alpha}(\omega)$
offers a possibility to track the evolution of remnant aspects of the 1D physics 
in the high-energy part of the spectrum.
Finally,  optical conductivity is a useful experimental~\cite{Basov05,Basov11} and 
numerical~\cite{Tohyama05,Lin09,Sedrakyan10,Lin11,Hartnoll11,Bergeron11,Chubukov14} 
response to study the interplay between the QP itineracy and the tendency towards their localization 
by dressing with AF spin fluctuations.

A real part of $\sigma_{\alpha}(\omega)$ is a measure of the 
rate at which particle-hole pairs are created by the absorption of photons with a given frequency $\omega$.
It might be calculated from the Kubo formula by looking at the QMC imaginary time 
current-current correlation functions,
\begin{equation}
\langle j_{\alpha}(\tau)  j_{\alpha}(0) \rangle  =
\frac{1}{\pi} \int {\rm d} w \;  \frac{e^{-\omega\tau}{\omega}}{1 - e^{-\beta \omega}} \; \sigma_{\alpha} (\omega).
\label{sigma}
\end{equation}
Here, $j_{\parallel}$  ($j_{\perp}$) is the intrachain (interchain) component of the current operator:
\begin{equation}
j_{\alpha} = i\sum_{\pmb{ij},\sigma}t^{}_{\pmb{ij}}\delta^{\alpha}_{\pmb{ij}}
   (c^{\dag}_{{\pmb i}\sigma}c^{}_{{\pmb j}\sigma} -h.c.),
\end{equation}
respectively, with $\delta^{\alpha}_{\pmb{ij}}$ being the $\alpha$ component of the 
vector connecting sites $\pmb i$ and $\pmb j$.

\subsubsection{Dimensional crossover}

\begin{figure}[t!]
\begin{center}
\includegraphics*[width=0.45\textwidth]{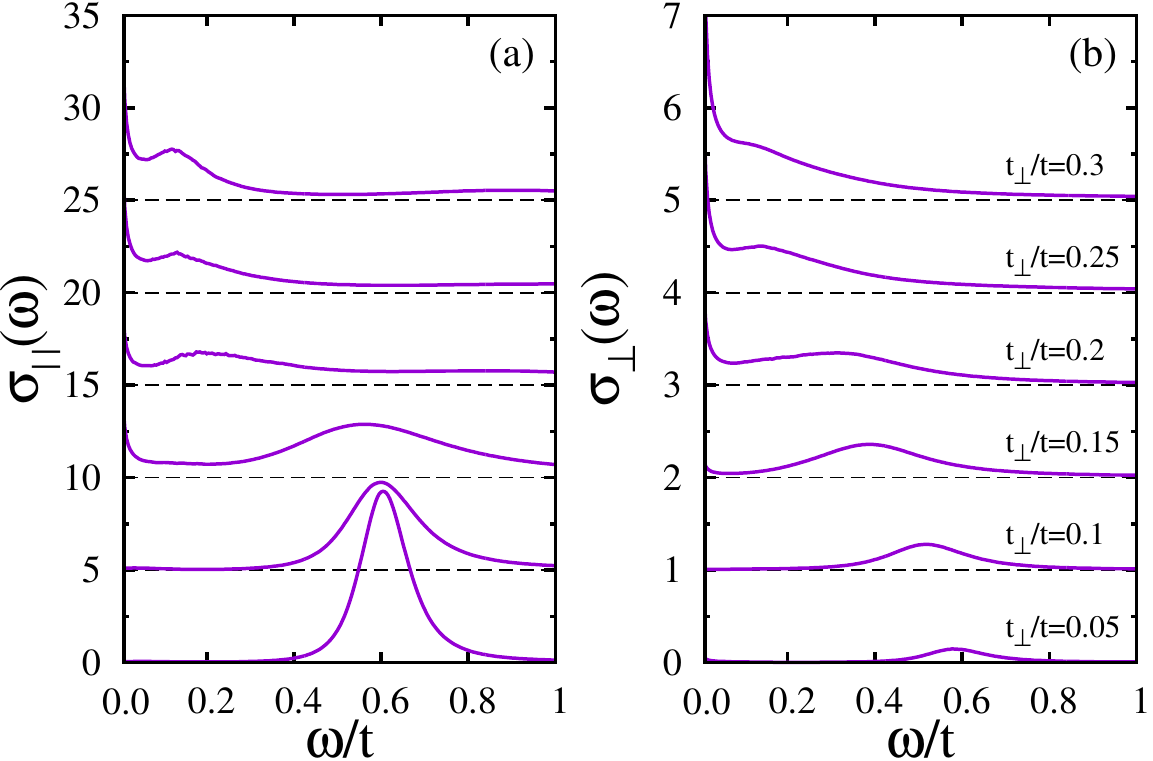}
\end{center}
\caption {
(Color online) Metal-insulator crossover in weakly coupled 1D chains upon increasing 
interchain hopping $t_{\perp}$. The plot shows: 
(a) intra- $\sigma_{\parallel}(\omega)$ and (b) interchain $\sigma_{\perp}(\omega)$  
optical spectra found on a $16\times 16$ lattice at $T=t/20$. 
From bottom to top: $t_{\perp}/t=0.05$, 0.1, 0.15, 0.2, 0.25, and 0.3.
}
\label{Sig_20}
\end{figure}

A dimensional-crossover-driven reconstruction of electronic states as evinced by 
frequency-dependent intra- $\sigma_{\parallel}(\omega)$ and interchain $\sigma_{\perp}(\omega)$ 
optical conductivities at a given temperature $T=t/20$ is shown in Fig.~\ref{Sig_20}.
At our smallest interchain coupling $t_{\perp}/t=0.05$, both optical conductivities 
display solely a finite-frequency feature typical of the 1D Mott insulating phase.~\cite{Jeckelmann00, Controzzi01} 
Its location $\Delta_{opt}/t\simeq 0.6$ roughly matches \emph{twice} the magnitude of 
the single-particle gap $\Delta_{qp}$, cf. Fig.~\ref{c00}, thus corresponding to the particle-hole pair 
absorption across the lower and upper Hubbard bands.    
While some smearing of this absorption mode becomes apparent,   
the peak retains its position in the weakly coupled regime with $t_{\perp}/t\le 0.1$. 
It reflects the relevance of umklapp processes and indicates charge confinement to the individual chains.

A further increase in $t_{\perp}$ renders umklapp scattering progressively irrelevant 
and the system enters a transient regime with a competing interchain single-particle tunneling. 
As a result, a Drude-like response in $\sigma_{\parallel}(\omega)$ develops  around $t_{\perp}/t\simeq 0.15$ 
accompanied by a tiny zero-frequency weight in $\sigma_{\perp}(\omega)$. Still, 
given a nearly $t_{\perp}$-independent position of the finite-frequency absorption,  
the overall structure of $\sigma_{\parallel}(\omega)$ is reminiscent of what is expected for a weakly 
doped 1D Mott insulator.~\cite{Giamarchi97,Carmelo00} 
Although there is no actual doping in the system, a non-negligible warping of the FS 
introduced by $t_{\perp}$ might be considered as an effective deviation from the commensurate filling 
of individual chains -- which otherwise -- continue to exhibit a substantial tendency to confine 
charge carriers reflected in a strongly reduced zero-frequency weight in $\sigma_{\perp}(\omega)$.

At larger coupling $t_{\perp}/t=0.2$, a pronounced Drude-like feature in $\sigma_{\perp}(\omega)$ 
signals the onset of a higher-dimensional metallic phase: 
single-particle \emph{interchain} tunneling requires the recombination of fractionalized 
excitations characteristic of the 1D regime into electronic QPs. 
The crossover in the effective dimensionality is not restricted to the lowest energies: 
a high-energy mode in $\sigma_{\parallel}(\omega)$, remnant of the 1D Mott gap, is replaced by 
a low-frequency hump whose evolution tracks the QP energy scale $\Delta_{qp}$.
The hump stems from coupling between charge carriers and short-range AF spin fluctuations 
beyond the 1D framework. A similar piling up of optical weight at finite frequency is also resolved 
in $\sigma_{\perp}(\omega)$ for $0.15\le t_{\perp}/t\le 0.25$ and then, in contrast to $\sigma_{\parallel}(\omega)$, 
the hump merges with the high-frequency tail of the Drude mode.
We attribute this difference to a spatial anisotropy in the spin-wave velocity;
the \emph{smaller} velocity associated with the \emph{interchain} spin-wave-like dispersion relation renders 
the paramagnon decay rate into particle-hole excitations \emph{higher} thus giving rise to 
a FL-like response in $\sigma_{\perp}(\omega)$.

\subsubsection{Thermal crossover}

\begin{figure}[t!]
\begin{center}
\includegraphics*[width=0.45\textwidth]{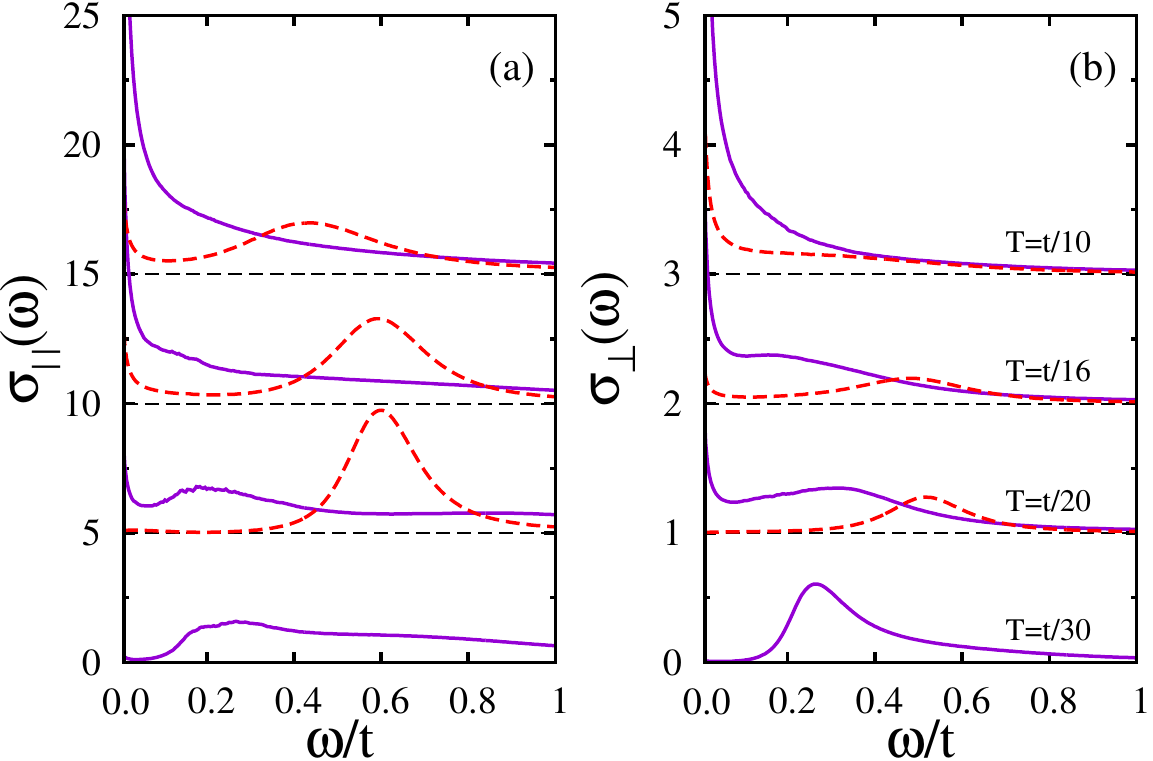}
\end{center}
\caption{
(Color online) Metal-insulator crossover in weakly coupled 1D chains upon decreasing
temperature $T$. The plot shows: 
(a) intra- $\sigma_{\parallel}(\omega)$ and (b) interchain $\sigma_{\perp}(\omega)$  
optical spectra at $t_{\perp}/t=0.2$ (solid) and $t_{\perp}/t=0.1$ (dashed).
From top to bottom: $T=t/10$, $t/16$, $t/20$, and $t/30$.
For $t_{\perp}/t=0.1$, a complete optical gap develops at $T=t/20$ and we expect 
at lower temperatures qualitatively similar spectra (not shown).
}
\label{Sig_T}
\end{figure}

Inspection of the static quantities in Sec.~\ref{Static} revealed the formation of a 
small QP gap $\Delta_{qp}$ associated with the onset of the  SDW state in the $T\to 0$ limit.
It is interesting to explore the behavior of this gap as a function of temperature. 
Thermal evolution of $\sigma_{\alpha}(\omega)$ at a fixed value 
of the interchain coupling $t_{\perp}/t=0.2$ is shown in Fig.~\ref{Sig_T}. 
At elevated temperature $T=t/10$, both intra- $\sigma_{\parallel}(\omega)$ and interchain $\sigma_{\perp}(\omega)$
optical conductivities exhibit broad Drude-like features exhausting most of the optical weight. 
This indicates a FL-like metal albeit with anisotropic transport properties due to 
inequivalent hopping amplitudes.  As the temperature is decreased down to $T=t/20$, 
a low-frequency depletion of optical weight separating the narrow Drude-like mode from the finite-frequency 
hump becomes apparent. It follows from a marked enhancement of the AF spin correlation length 
and dressing of QPs with a cloud of spin fluctuations. 
Upon further cooling, the AF spin correlations extend across the entire lattice and the 
SDW gap opens up.

We now compare the above evolution of $\sigma_{\alpha}(\omega)$
with the corresponding redistribution of optical weight at a twice \emph{smaller} interchain coupling $t_{\perp}/t=0.1$.    
The low-temperature increase of the static spin structure factor $S(\pi,0)$ reveals 
dominant 1D spin correlations in this regime, see Fig.~\ref{S_T}(a). 
A pronounced 1D character of the system is also reflected in a 
different frequency dependence of intra-  and interchain optical conductivities. 
This effect is particularly dramatic at $T=t/10$: 
while a clear finite-frequency depletion of optical weight is visible in $\sigma_{\parallel}(\omega)$, 
only a narrow Drude-like peak is resolved in $\sigma_{\perp}(\omega)$. 
In spite of thermal broadening comparable with the interchain bandwidth
masking partially warping of the FS, the presence of the zero-frequency mode in 
$\sigma_{\perp}(\omega)$ suggests that charge carriers retain their electronic QP nature.  
This indicates that thermal melting of the quasi-1D Mott insulator induces at the \emph{smallest} energy 
scales a crossover to the FL characteristics rather than to the LL behavior.
Furthermore, at the intermediate low temperature $T=t/16$, the frequency range with a reduced
weight in $\sigma_{\parallel}(\omega)$ becomes wider thus signaling proximity to the insulating phase. 
A reduced mobility of charge carriers also affects $\sigma_{\perp}(\omega)$; here,         
most of the Drude-like weight is transferred to a high-energy feature. 
Finally, a complete optical gap driven by umklapp scattering develops already at $T=t/20$.

\subsection{Single-particle spectral function}

Our next aim is to address momentum-resolved single-particle spectral properties.
In particular, one would like to know (i) whether the quasi-1D metallic phase can be 
described within the FL theory, and (ii) topology and QP weight along the emergent FS.  
Another interesting question is whether some spectral features in the higher-dimensional 
electronic structure can be traced back to those of the 1D regime.~\cite{Kohno12,Kohno14}

To gain insight into these issues, we compute the momentum-resolved single-particle spectral 
function $A({\pmb k},\omega)$ which is related to the QMC imaginary time Green's function 
$G({\pmb k},\tau)$ by the spectral theorem:  
\begin{equation}
G({\pmb k},\tau)=
\frac{1}{\pi} \int {\rm d} w \;  \frac{e^{-\omega\tau}}{1 + e^{-\beta \omega}} \; A({\pmb k},\omega).
\end{equation}
\begin{figure}[t!]
\begin{center}
\includegraphics*[width=0.43\textwidth]{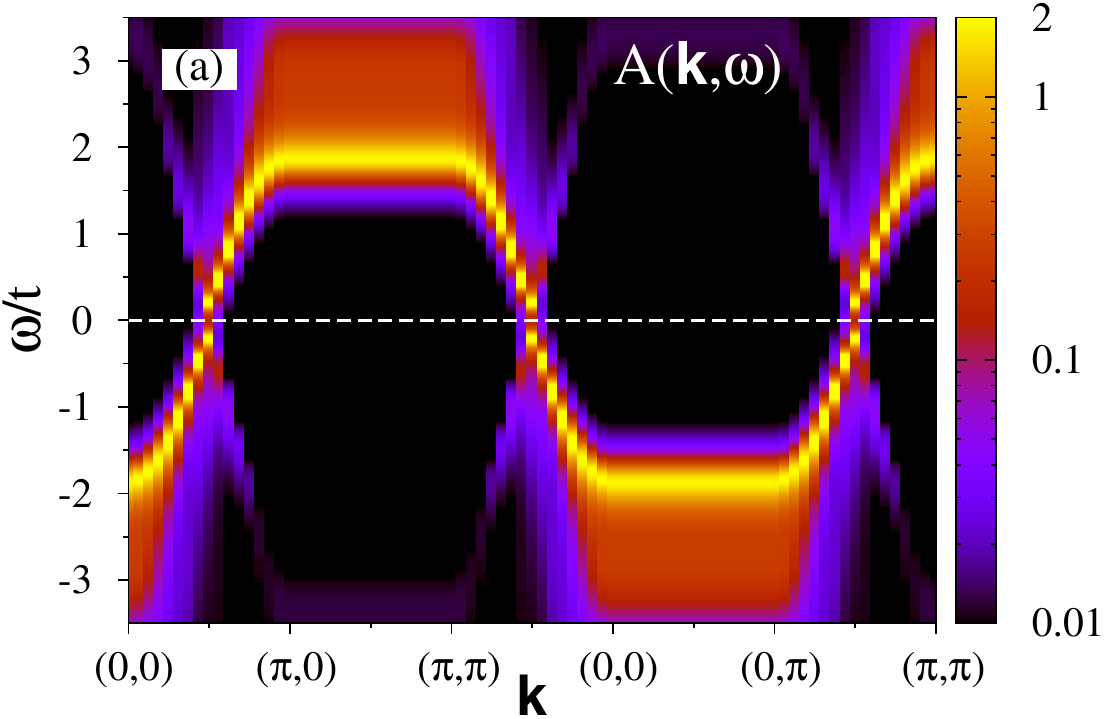}\\
\includegraphics*[width=0.43\textwidth]{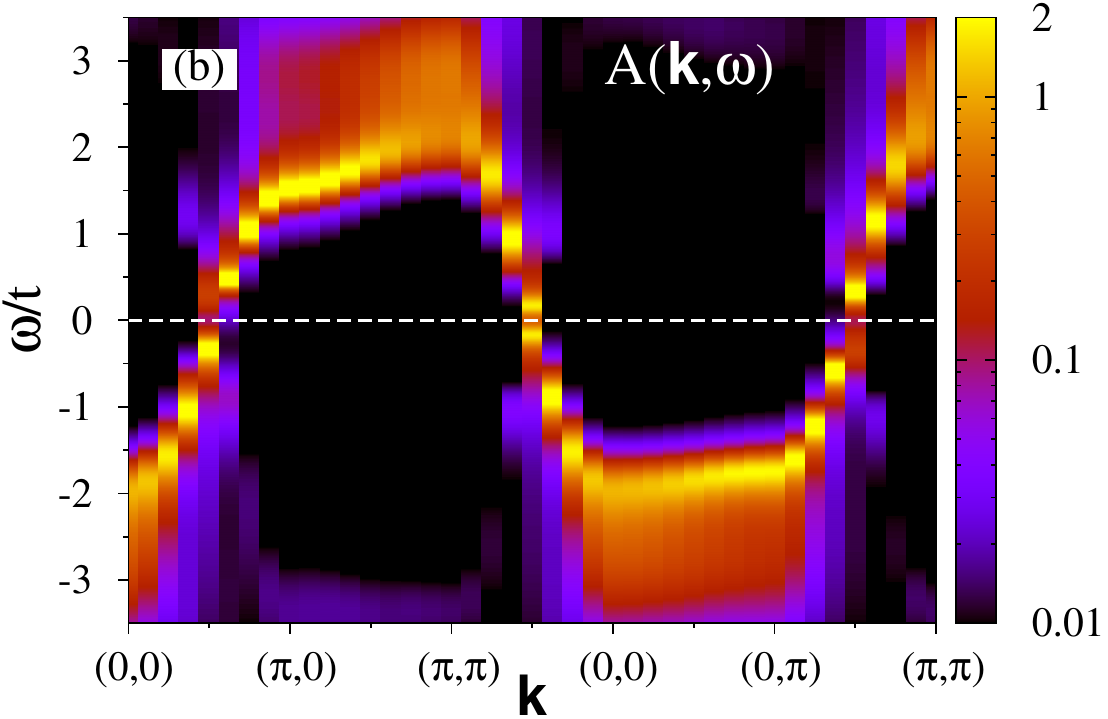}\\
\includegraphics*[width=0.43\textwidth]{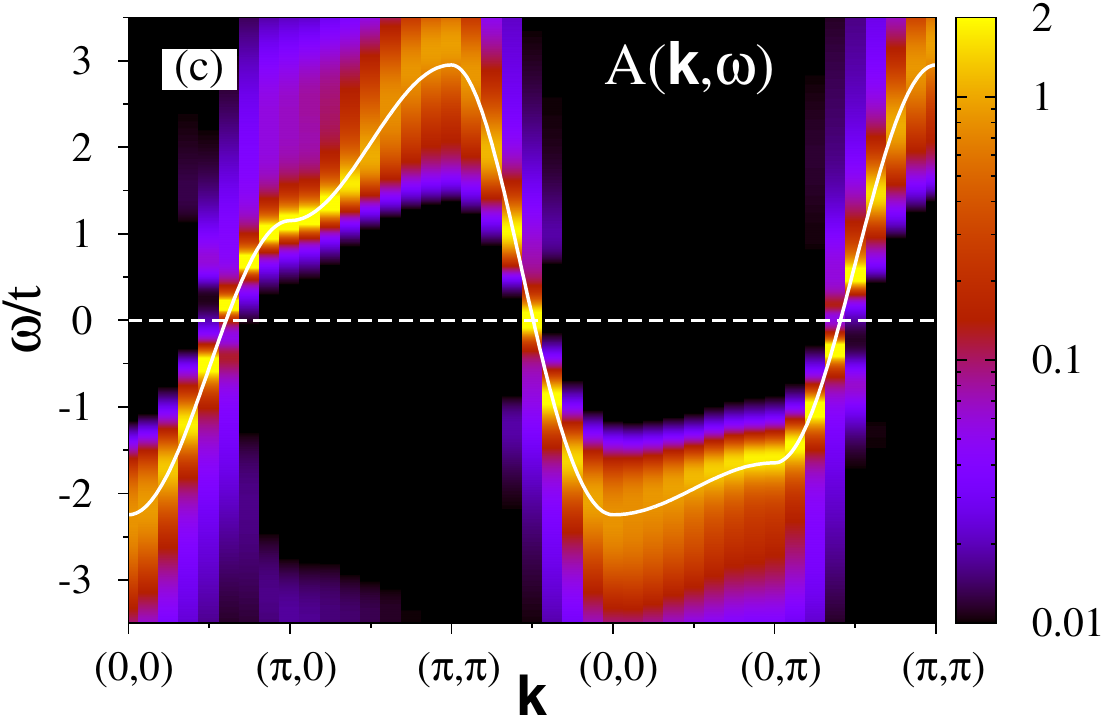}
\end{center}
\caption {(Color online) Single-particle spectral function $A({\pmb k},\omega)$ obtained 
at $T=t/20$ on: 
(a) 32-site chain with a dummy transverse momentum and (b),(c) $16\times 16$ lattice with: 
$t_{\perp}/t=0.15$ (middle) and $t_{\perp}/t=0.3$ (bottom). 
Solid line in panel (c) gives the paramagnetic Hartee-Fock band structure.
}
\label{Akw}
\end{figure}
The evolution of $A({\pmb k},\omega)$ with increasing interchain coupling $t_{\perp}$ is summarized 
in Fig.~\ref{Akw}. In the 1D regime, the relevance of umklapp process generates a gap in the half-filled 
band at ${k}_{\rm F}=\pm\pi/2$ and the equivalent points, see Fig.~\ref{Akw}(a). 
At frequencies above the single-particle gap, the dispersion is approximately linear    
reflecting aspects of the LL theory.~\cite{Benthien04,Abend06} 
Moreover, the spectrum significantly broadens on approaching the Brillouin zone edges reminiscent of the spinon and holon 
branches.~\cite{Preuss94,Senechal00,Aichhorn04,Matsueda05}
Next, Fig.~\ref{Akw}(b) provides evidence for highly incoherent single-particle dynamics in 
the metallic regime with $t_{\perp}/t=0.15$; a significant depletion of spectral weight at 
the ${\pmb k}=(\pi/2,\pi/2)$ momentum is accompanied by a backfolding behavior in the dispersion 
at ${\pmb k}=(\pi/2,0)$ and ${\pmb k}=(\pi/2,\pi)$ points. 
Upon further increase in the interchain coupling strength, the spectral intensity 
at the ${\pmb k}=(\pi/2,\pi/2)$ point gradually recovers. 
Finally, a single QP peak becomes apparent at the Fermi level at $t_{\perp}/t=0.3$ and 
the low-frequency part of $A({\pmb k},\omega)$ follows roughly the paramagnetic Hartee-Fock band structure, 
see Fig.~\ref{Akw}(c).

Another noteworthy feature is found in the inverse photoemission $\omega>0$ part of the spectra, 
i.e., the formation of a weakly dispersive QP-like band near the ${\pmb k}=(\pi,0)$ momentum.  
While the flatness is reminiscent of the 1D nature of the problem, 
only a broad structure is resolved in the photoemission $\omega<0$ part around the ${\pmb k}=(0,\pi)$ point. 
The difference between the inverse- and photoemission parts reflects a broken 
particle-hole symmetry due to the finite next-nearest-neighbor hopping $t'$.

\begin{figure}[t!]
\begin{center}
\includegraphics[width=0.35\textwidth]{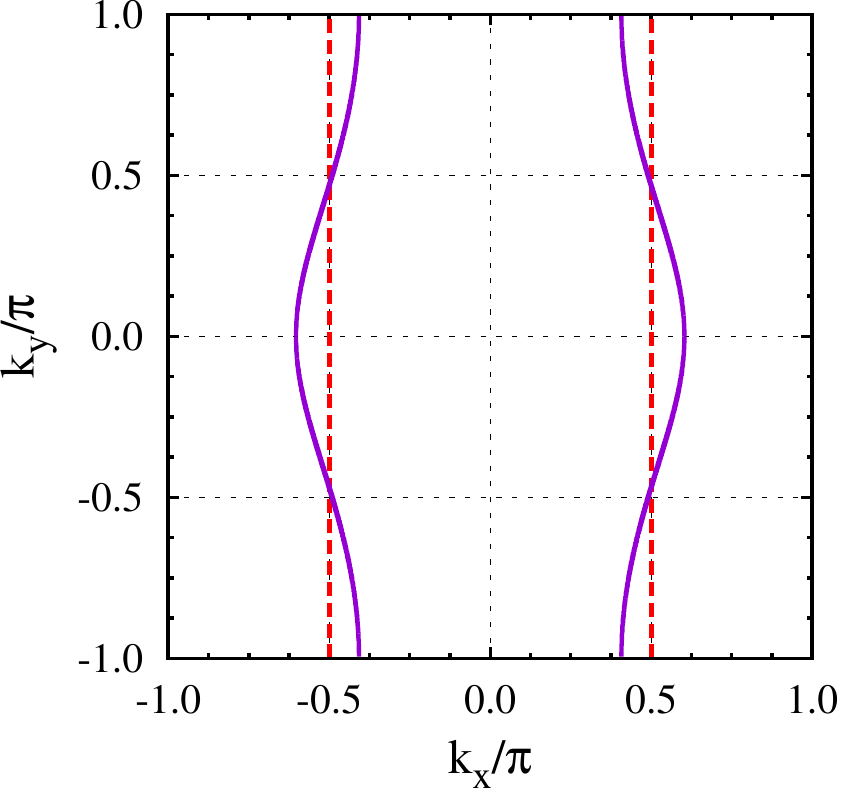}
\end{center}
\caption {(color online)
The noninteracting Fermi surface with $t_{\perp}/t=0.3$ (solid) and of the purely 1D case (dashed).}
\label{FS}
\end{figure}

The evolution of the spectral function $A({\pmb k},\omega)$ is a consequence of dramatic 
changes in the single-particle Green's function $G({\pmb k},\omega)$.  
In the Mott insulator, the presence of a spectral gap requires that the real part 
of the zero-frequency Green's function $\textrm{Re}\, G({\pmb k},\omega)$ 
changes sign in momentum space by going through a \emph{zero}. 
This is accomplished by the singularity in the corresponding self-energy. 
As one approaches the Mott transition, the locus of zeros in ${\pmb k}$-space 
affects the topology of the emergent FS defined by zero-frequency \emph{poles} 
of the Green's function.~\cite{Stanescu06,Stanescu07,Sakai09,Tocchio12} 
We address this issue by examining two special momenta: 
(i) nodal ${\pmb k}=(\pi/2,\pi/2)$, corresponding to a vanishing interchain kinetic 
energy $\partial \varepsilon_{\pmb k}/\partial t_{\perp} = 0$, 
and (ii) an antinodal ${\pmb k}=(\pi/2,0)$ one, where the maximum warping of the 
noninteracting 1D FS occurs, see Fig.~\ref{FS}.

\begin{figure}[t!]
\begin{center}
\includegraphics[width=0.45\textwidth]{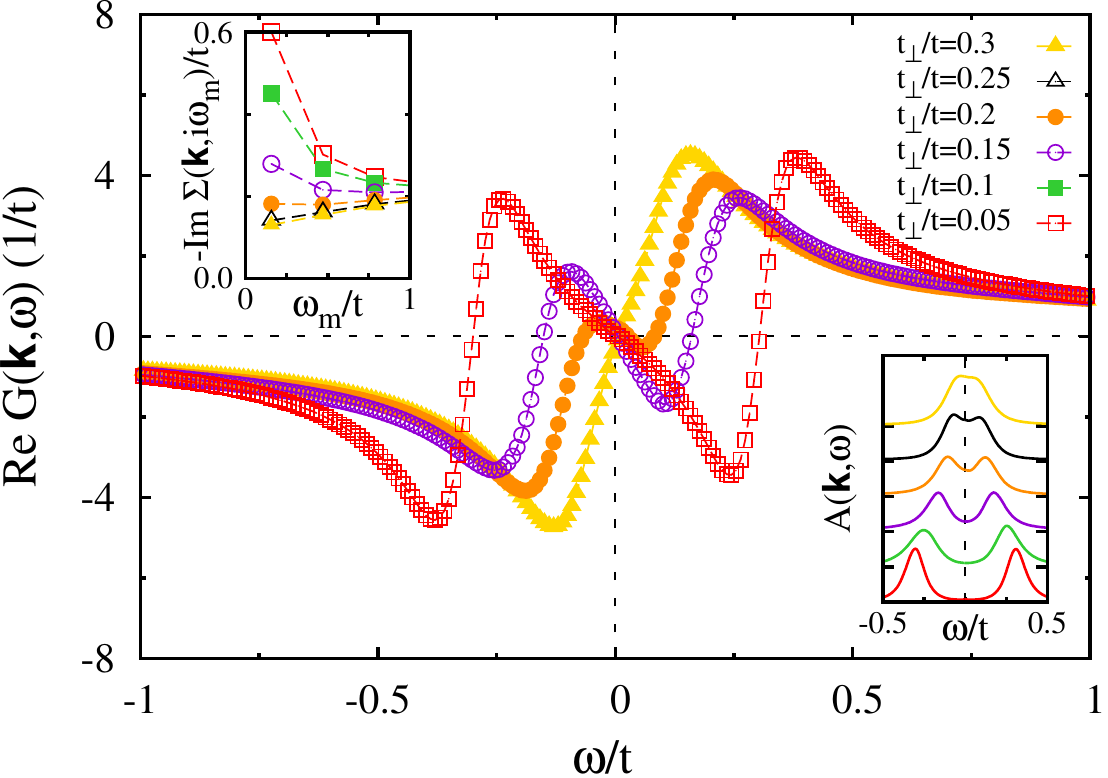}
\end{center}
\caption {(color online)
Dimensional-crossover-driven evolution of the real part of the Green's function for  fixed $T=t/20$ 
at the nodal ${\pmb k}=(\pi/2,\pi/2)$ point. 
Insets show the corresponding: 
(left) low-frequency dependence of the imaginary part of the self-energy 
and (right) single-particle spectral function 
from bottom to top: $t_{\perp}/t=0.05$, 0.1, 0.15, 0.2, 0.25, and 0.3. 
}
\label{G_nod}
\end{figure}

We focus first on the nodal ${\pmb k}=(\pi/2,\pi/2)$ point considered in Fig.~\ref{G_nod}. 
At our smallest $t_{\perp}/t=0.05$, $\textrm{Re}\, G({\pmb k},\omega)$ has a negative slope 
in a broad range of frequencies around the Fermi level. 
The imaginary part of the corresponding self-energy $-\textrm{Im}\, \Sigma({\pmb k},i\omega_m)$ 
displays a diverging-like behavior on approaching the smallest Matsubara frequency $\omega_0=\pi T$ 
thus signaling a zero of $\textrm{Re}\, G({\pmb k},\omega)$, see Fig.~\ref{G_nod}, left inset. 
The anomalous behavior of the self-energy stems from umklapp scattering and is responsible for a robust Mott gap 
in the single-particle spectral function $A({\pmb k},\omega)$, right inset of Fig.~\ref{G_nod}. 
At larger $t_{\perp}$, the umklapp process becomes less effective at low-energy scales. 
This shrinks the frequency region with a negative slope of $\textrm{Re}\, G({\pmb k},\omega)$ 
and reduces the scattering rate $\Gamma_{\pmb k}=-\textrm{Im}\, \Sigma({\pmb k},\omega_0)$.  
As a result, some thermally excited single-particle states whose weight is controlled by $t_{\perp}$ 
become apparent at the Fermi energy.  
Finally, $\textrm{Re}\, G({\pmb k},\omega)$ develops a positive slope at $t_{\perp}/t=0.3$ thus forming a 
pole-like structure as in the FL phase. Still, a small kink at $\omega=0$  signals substantial 
QP scattering off AF spin fluctuations. Consequently, a broad QP-like feature is resolved 
in $A({\pmb k},\omega)$.

\begin{figure}[t!]
\begin{center}
\includegraphics[width=0.45\textwidth]{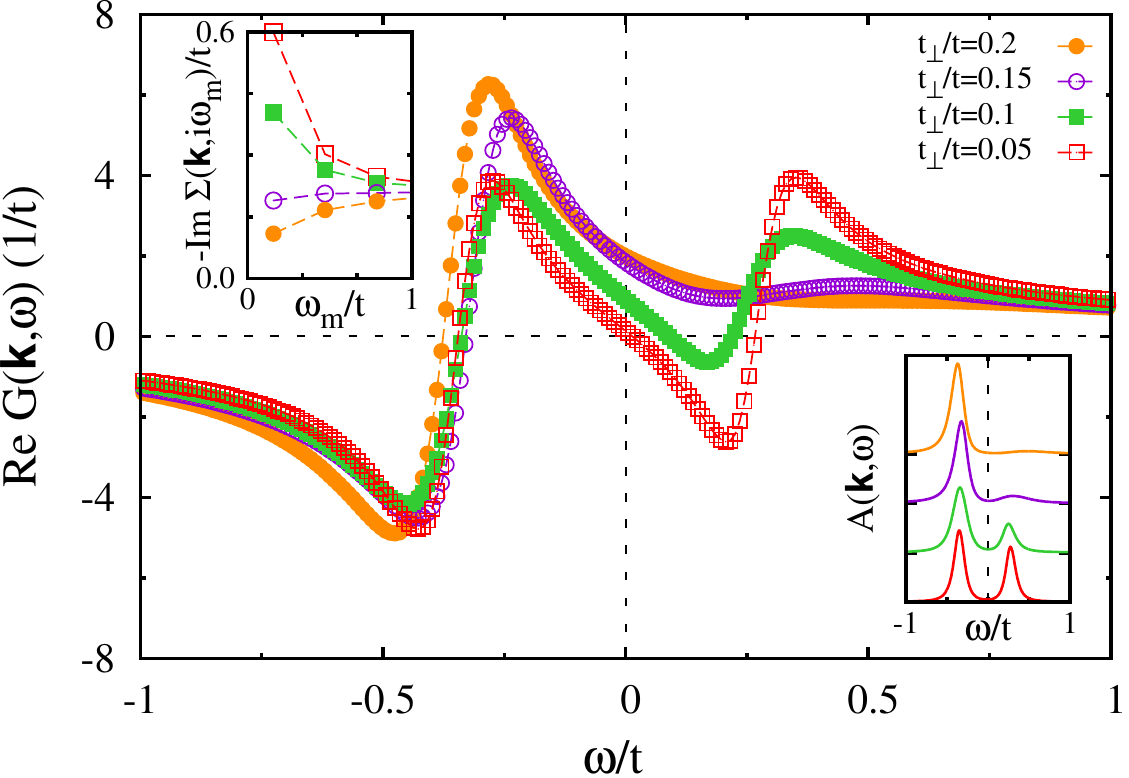}
\end{center}
\caption {(color online)
Same as in Fig.~\ref{G_nod} but at the antinodal ${\pmb k}=(\pi/2,0)$ point. 
}
\label{G_anod}
\end{figure}

We turn now to the antinodal ${\pmb k}=(\pi/2,0)$ point.
As shown in Fig.~\ref{G_anod}, the zero of $\textrm{Re}\, G({\pmb k},\omega)$ and 
the Mott gap in the spectral function $A({\pmb k},\omega)$ remain for small values 
of $t_{\perp}/t<0.1$ nearly pinned at the ${\pmb k}=(\pi/2,0)$ momentum. 
Hence, at the expense of loss in the \emph{interchain} kinetic energy, 
the interaction renders the FS warping tendency irrelevant.~\cite{Dusuel03}
In contrast, at larger interchain hopping amplitude, a reduced scattering rate  
$\Gamma_{\pmb k}$ indicates that the kinetic energy gain 
cannot be further ignored and the warping effects become discernible. Indeed, 
vanishing $\textrm{Re}\, G({\pmb k},\omega)=0$ at the ${\pmb k}=(\pi/2,0)$ point
requires now a finite frequency $\omega>0$ thus approaching a pole-like behavior around $\omega/t=0.2$. 
The latter produces a faint spectral feature in $A({\pmb k},\omega)$; 
it signals backfolding of the conduction band and as such is a fingerprint 
of the Mott gap slightly off the ${\pmb k}=(\pi/2,0)$ momentum.  
Finally, the finite-frequency zero and pole-like features annihilate together 
and only a shallow dip in $\textrm{Re}\, G({\pmb k},\omega)$ is resolved at $t_{\perp}/t=0.15$ 
which in turn evolves into a smooth behavior at $t_{\perp}/t=0.2$.   

Obviously, we cannot resolve the Fermi wave vector ${\pmb k}_{\rm F}$ 
across the $(0,0)\to (\pi,0)$ path on our discrete ${\pmb k}$-point mesh.
Given, however, a comparable slope of the QP energy dispersion  
near the antinodal ${\pmb k}=(\pi/2,0)$ and nodal ${\pmb k}=(\pi/2,\pi/2)$ points, 
see Fig.~\ref{Akw}(c), we expect a \emph{continuous} warped FS. 
This should be contrasted with the isotropic two-dimensional Hubbard model at half-filling where 
the flatness of the QP band around the ${\pmb k}=(\pi,0)$ point renormalizes locally 
the Fermi velocity and enhances QP scattering off AF spin fluctuations. 
As a result, the depletion of QP weight starts at different temperatures 
in different regions of the Brillouin zone.~\cite{Rost12} 
However, this momentum-sector-selective opening of the single-particle spectral gap is suppressed by 
next-nearest-neighbor hopping $t'$.~\cite{Gull09} 
The latter shifts the flat region in the QP dispersion away from the Fermi level thus providing further support for 
the continuous FS in the quasi-1D regime.

The evolution of the QMC Green's function $G({\pmb k},\omega)$  and the shape of the emergent FS are different 
from those concluded on the basis of the CDMFT approach on the $8\times 2$ cluster.~\cite{Raczkowski12} 
In CDMFT, the range of AF spin fluctuations is restricted to the cluster size. Therefore, 
the evolution of the electronic structure is governed by dominant umklapp process:
a finite $t_{\perp}$ shifts the \emph{zero} of $\textrm{Re}\, G({\pmb k},\omega)$ away from 
the $(\pm\pi/2,0)$ points giving way to a finite-frequency \emph{pole} 
from the conduction band, see Fig. 5(a) in Ref.~\onlinecite{Raczkowski12}. 
At a critical value of $t_{\perp}$, the latter crosses the Fermi level which opens up  
elongated \emph{electron} pockets centered around the ${\pmb k}=(\pm\pi/2,0)$ momenta.
At the same time, \emph{hole} pockets open up at the ${\pmb k}=(\pm\pi/2,\pi)$ points 
and thus a \emph{broken} FS emerges with the 1D nesting wave vector ${\pmb k}=(\pi,0)$. 
Hence, in the absence of finite-range AF spin fluctuations, this particular 
evolution of $\textrm{Re}\, G({\pmb k},\omega)$ ensures the relevance of  umklapp 
scattering and extends the stability of the Mott phase away from the 1D regime.
Let us however point out that a larger next-nearest-neighbor hopping $t'$  in the lattice calculations  
will frustrate the AF spin correlations and could bring our findings in line with the CDMFT results. 
Unfortunately, this leads to a sign problem which renders  QMC simulations very expensive.

\subsection{Dynamical spin and charge structure factors}

Competing effects in strongly correlated systems often result in various energy scales 
which govern the population of  excited states. 
This also holds for weakly coupled 1D Hubbard chains with a half-filled band:  
in the regime where the interaction between spin and electronic degrees of freedom 
becomes relevant, both single- and two-particle processes contribute to the excited state dynamics. 
The energy- and momentum-resolved spin and charge structure factors 
help to separate these contributions which is crucial for the understanding 
of the low-energy dynamics. 
In addition, these quantities allow one to identify momentum-resolved energy scales 
above which spin and charge fluctuations lose their 1D nature.  

Thus, to complement the characteristic of the emergent quasi-1D metallic phase,  we consider 
the spin $S({\pmb q},\omega)$  and charge  $C({\pmb q},\omega)$  dynamical structure factors 
defined as: 
\begin{equation} 
	S({\pmb q},\omega) = \frac{\chi_s''({\pmb q},\omega)}{1-e^{-\beta \omega}} 
\; \; {\rm and } \; \;
       C({\pmb q},\omega) = \frac{\chi_c''({\pmb q},\omega)}{1-e^{-\beta \omega}}. 
\end{equation}
Here,  $\chi_c$ and $\chi_s$ correspond to the generalized  charge and spin susceptibilities. 
The susceptibilities  can be obtained from the imaginary-time displaced two-particle correlation functions,
\begin{align}
\langle S^z({\pmb q}, \tau)  S^z(-{\pmb q},0) \rangle  &= 
\frac{1}{\pi} \int {\rm d} w \;  \frac{e^{-\omega\tau}}{1 - e^{-\beta \omega}} \; \chi_s'' ({\pmb q},\omega),  \label{sqw} \\ 
\langle N({\pmb q}, \tau)   N(-{\pmb q},0) \rangle     &= 
\frac{1}{\pi} \int {\rm d} w \; \frac{e^{-\omega\tau}}{ 1 - e^{-\beta \omega} } \;  \chi_c''({\pmb q},\omega) \label{cqw},
\end{align}
where,
\begin{align} 
 S^z({\pmb q})  &= \frac{1}{\sqrt{N} } \sum_{\pmb r} e^{i {\pmb q}\cdot {\pmb r} } 
\left( n_{{\pmb r}\uparrow} - n_{{\pmb r}\downarrow}  \right),  \label{sz}\\ 
  N({\pmb q})  &= \frac{1}{\sqrt{N} } \sum_{\pmb r} e^{i {\pmb q}\cdot {\pmb  r} } 
\left( n_{{\pmb r}\uparrow} + n_{{\pmb r}\downarrow}  - n\right) \label{n}, 
\end{align}
with $n=\sum_{\sigma}\langle n_{{\pmb r},\sigma}\rangle$ being the average filling level.

\subsubsection{1D Mott insulator}

\begin{figure}[t!]
\begin{center}
\includegraphics*[width=0.43\textwidth]{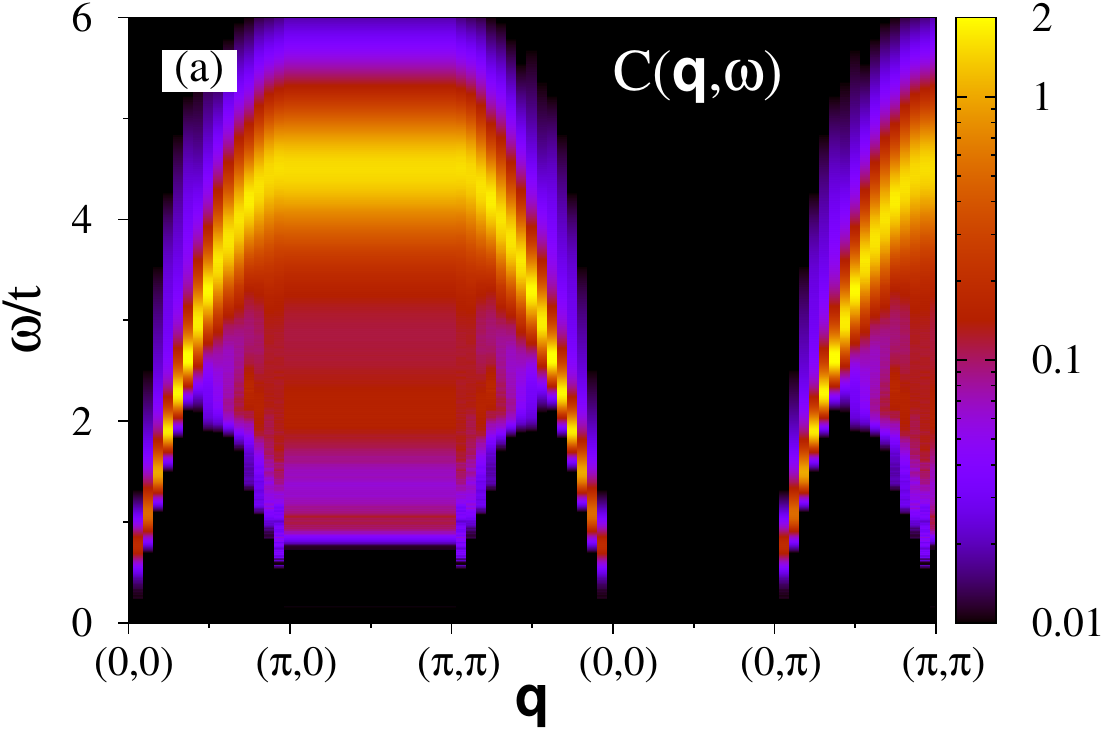}\\
\includegraphics*[width=0.43\textwidth]{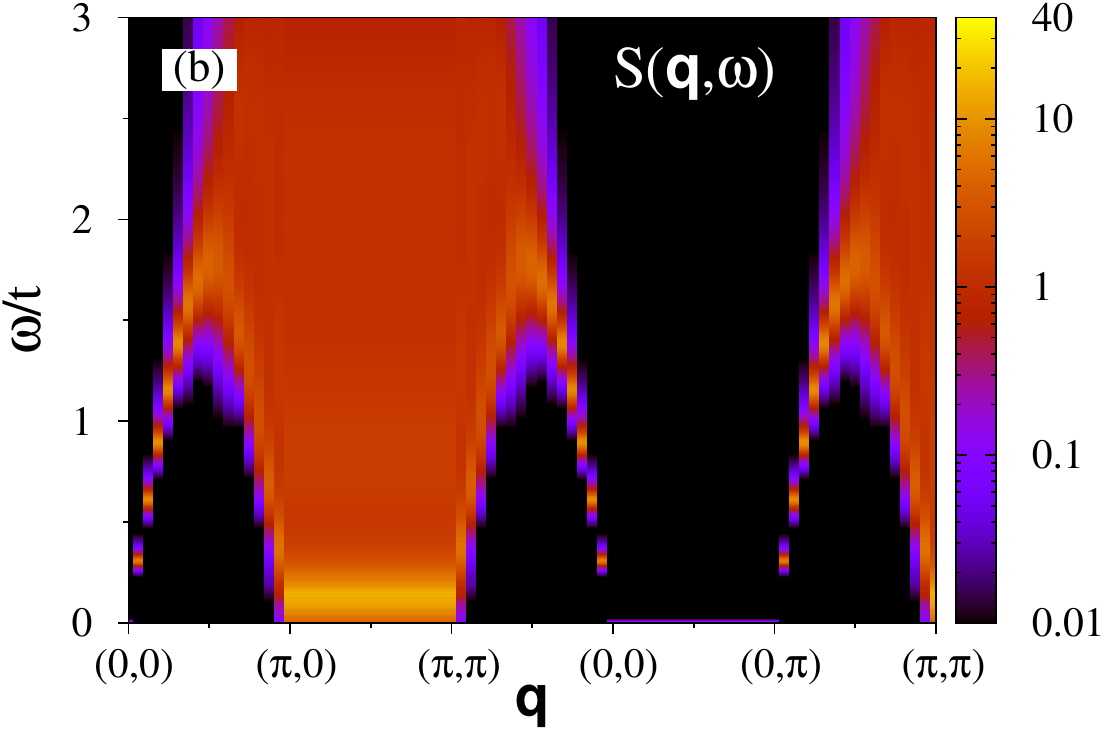}\\
\end{center}
\caption {(Color online) (a) Dynamical charge structure factor $C({\pmb q},\omega)$ 
 and (b) dynamical spin structure factor $S({\pmb q},\omega)$ obtained on a 32-site chain at $T=t/20$. 
The transverse momentum is a dummy label.
}
\label{chain}
\end{figure}

The elementary excitations of the 1D Hubbard model are bosonic collective spin and charge 
density oscillations with a linear dispersion in the long-wavelength limit ${\pmb q}\to 0$.~\cite{Giamarchi_book}
The relevance of umklapp scattering at half-filling opens up a gap for  long-wavelength 
charge excitations while leaving the spin sector gapless, see Fig.~\ref{chain}.  
The dynamical charge structure factor $C({\pmb q},\omega)$ in Fig.~\ref{chain}(a) shows an overall agreement  
with the one obtained at a slightly smaller interaction $U/t=2$ using the time-dependent density 
matrix renormalization group method in the $T=0$ limit, see Fig. 10 in  Ref.~\onlinecite{Penc12}. 
Above the charge gap, one finds aspects of the LL dynamics with low-lying charge excitations 
located at long wavelengths as well as at ${\pmb q} = 2 k_{\rm F} = \pi$. As found in Ref.~\onlinecite{Penc12},  
the latter carry little spectral weight and are thus more difficult to resolve in our 
finite-temperature QMC simulations.       

Since the charge sector is fully gapped, the spin dynamics can be understood within a spin-only $S=1/2$ 
Heisenberg chain. In this case, the spin dynamics is characterized by the two-spinon continuum 
of states bounded from below and above by,~\cite{Clo62}
\begin{equation}
\frac{\pi}{2}J|\sin({\pmb q})|\leq\omega({\pmb q})\leq\pi J|\sin({\pmb q}/2)|,
\label{CP}
\end{equation}
with the lowest-lying excitations carrying dominant spectral weight
at low temperatures.~\cite{Jarrell93,Sandvik97,Benthien07,Barthel09,Lake13,Ronnow13}
As is apparent in Fig.~\ref{chain}(b), for the considered value of the interaction $U/t=2.3$, 
the majority of spectral weight is located at the lower bound of the continuum 
as in the Heisenberg model.

In spatial dimensions greater than one, the spinons bind together and form magnons, 
the Goldstone modes of the broken continuous SU(2) symmetry group.    
As a result, the continuum of excitations in the dynamical spin structure factor
gives way to well defined spin-wave modes.~\cite{Raczkowski13} 
The low-frequency magnons in AF \emph{insulators} and in the $T=0$ limit are well accounted for 
within the linear spin-wave theory (LSWT) of the Heisenberg model to leading $1/S$ order.~\cite{Anderson52,Kubo52,Raczkowski02}
This picture should break down at finite temperatures where 
cranking up the interchain coupling triggers the crossover to a \emph{metal}.

\subsubsection{Quasi-1D metal in the pseudogap regime}

Let us consider first  a weakly coupled regime with $t_{\perp}/t=0.15$. 
As shown in Fig.~\ref{G_nod}, the corresponding single-particle spectral function 
$A({\pmb k},\omega)$  exhibits two broadened peaks separated by a region with a strongly reduced weight that we 
refer to as a \emph{pseudogap}. As is apparent, the system is at the verge of localization near the 1D Mott phase.
 
\begin{figure}[t!]
\begin{center}
\includegraphics*[width=0.43\textwidth]{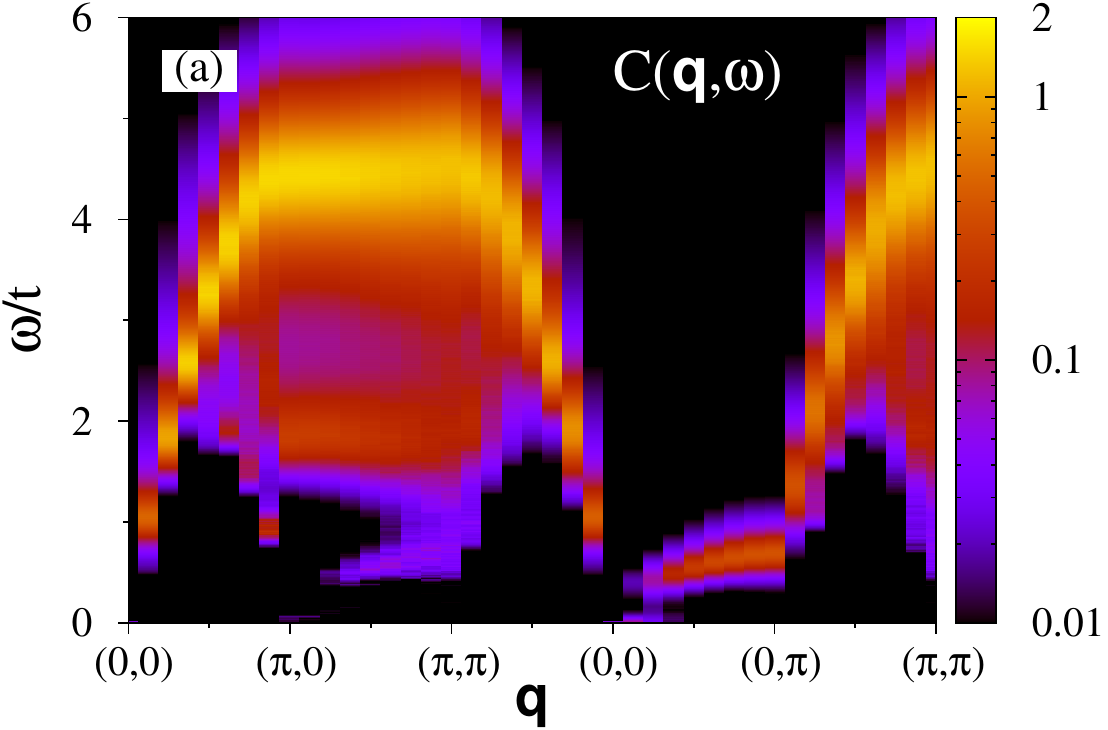}\\
\includegraphics*[width=0.43\textwidth]{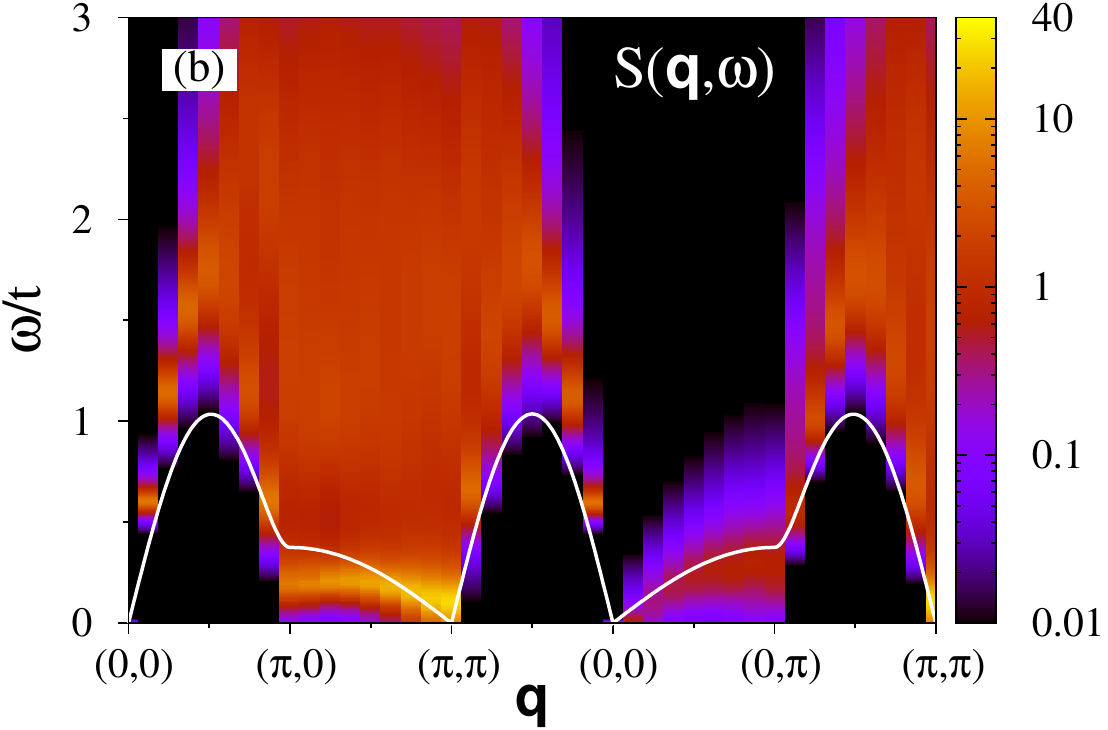}\\
\end{center}
\caption {(Color online) (a) Dynamical charge structure factor $C({\pmb q},\omega)$, 
and (b) dynamical spin structure factor $S({\pmb q},\omega)$ 
obtained on a $16\times 16$ lattice with $t_{\perp}/t=0.15$ at $T=t/20$. 
Solid line in panel (b) gives the LSWT dispersion Eq.~(\ref{magnon}) 
with $J_{\perp}/J=0.06$ and $J'/J_{\perp}=0.2$. 
The chosen exchange couplings are fit parameters subject to the constraint of a finite 
magnetic order parameter in LSWT consistent with the long-range AF order in the system.
}
\label{dyn15}
\end{figure}

Figure~\ref{dyn15}(a) depicts the dynamical charge structure factor $C({\pmb q},\omega)$. 
Using the continuity equation one finds that upon approaching the 
long-wavelength limit ${\pmb q}\to 0$ parallel (perpendicular) to the chains,
this observable is related to the intrachain (interchain) real part of the optical conductivity 
$\sigma_{\alpha}(\omega)$, respectively:~\cite{Assaad08}
\begin{equation}
\label{Sigma_Cqw}
        \sigma_{\alpha}(\omega)   = 
\lim_{{q_{\alpha}} \to 0}\frac{\omega}{ {q}^2_{\alpha} } 
        \left( 1 - e^{-\beta \omega} \right) C( q_{\alpha},\omega),
\end{equation}
where $q_{\alpha}$ is the $\alpha$ component of the momentum transfer parallel/perpendicular to the chains. 
Given that the intensity of charge fluctuations rapidly decreases in the long-wavelength 
limit, it is difficult to pin down the metallic vs. insulating nature of the system. 
This issue is resolved by looking at the low-frequency part of the optical conductivity 
$\sigma_{\alpha}(\omega)$, see Fig.~\ref{Sig_20}.
On the one hand, zero-frequency weight in the interchain optical conductivity $\sigma_{\perp}(\omega)$ indicates 
a metallic phase at $t_{\perp}/t=0.15$. 
On the other hand, the smallness of this weight reveals a large degree of incoherence in the interchain transport. 
Moreover, as shown in Fig.~\ref{G_nod}, the imaginary part of the corresponding self-energy   
$-\textrm{Im}\, \Sigma({\pmb k},\omega_m)$ increases at low Matsubara frequency 
contributing to the effective mass enhancement and \emph{reduced} mobility of charge carriers.

From this standpoint, we can analyze the magnetic excitation spectrum shown in Fig.~\ref{dyn15}(b).
The most striking difference with respect to the 1D regime is a low-frequency dispersive  
feature  along the $(\pi,\pi)\to (\pi,0)$ path accompanied by a broad upward dispersion 
along the $(0,0)\to(0,\pi)$ direction. 
They are signatures of damped AF spin fluctuations (paramagnons) which are not strong enough to 
gap out the FS and  develop long-range AF order but nevertheless can propagate an appreciable distance. 

To get  further insight into the spin dynamics, we consider the spin $S=1/2$ Heisenberg model 
with nearest neighbor interactions $J$ ($J_{\perp}$) along the intrachain (interchain) bonds, 
respectively, extended by next-nearest neighbor interaction $J'$:
\begin{equation}
H_{J}=
J\sum_{\langle\pmb{ij}\rangle_{\parallel}} {\pmb S}_{\pmb i}\cdot{\pmb S}_{\pmb j}
  +J_{\perp}\sum_{\langle\pmb{ij}\rangle_{\perp}} {\pmb S}_{\pmb i}\cdot{\pmb S}_{\pmb j}
  +J'\sum_{\langle\langle\pmb{ij}\rangle\rangle } {\pmb S}_{\pmb i}\cdot{\pmb S}_{\pmb j}.
\label{HJ}
\end{equation}
As illustrated in Fig.~\ref{dyn15}(b), the LSWT dispersion relation of the Heisenberg 
model (\ref{HJ}),~\cite{Raczkowski13} 
\begin{equation}
\omega_{\pmb q}=2S\sqrt{ \xi^2_{\pmb q} -\gamma^2_{\pmb q} },
\label{magnon}
\end{equation}
with two-dimensional structure factors,
\begin{align}
\xi_{\pmb q} &= J+J_{\perp}-2J'(1-\cos q_{x}\cos q_{y}),\\
\gamma_{\pmb q} &= J\cos q_{x}+J_{\perp}\cos q_{y},
\end{align}
tracks the low-frequency paramagnon spectrum.
Clearly, the frequency region with  depleted single-particle spectral weight 
efficiently separates the low-frequency spin-wave-like excitations from the high-frequency 
particle-hole excitations across the pseudogap in $A({\pmb k},\omega)$. 
Moreover, a reduced mobility of charge carriers leads to a separation of time 
scales, a characteristic feature of correlated systems with complex dynamics;~\cite{Raczkowski10} 
charge carriers are \emph{localized} on the spin time scale and thus 
a coherent-like precession of individual spins might still take place.
Consequently, despite broadening and renormalization of the paramagnon dispersion (softening)                 
as compared to LSWT with localized moments,
an approximate Heisenberg-like picture turns out to still be applicable below 
the single-particle pseudogap.  

\subsubsection{Quasi-1D metal with quasiparticles }

\begin{figure}[t!]
\begin{center}
\includegraphics*[width=0.43\textwidth]{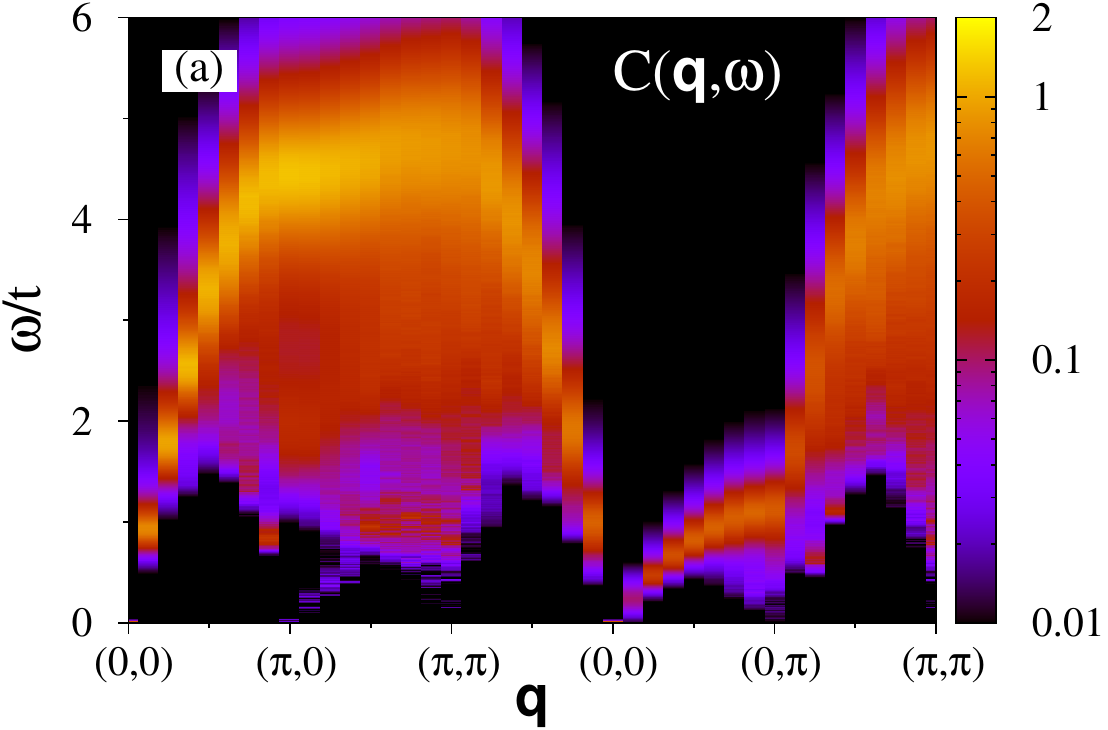}\\
\includegraphics*[width=0.43\textwidth]{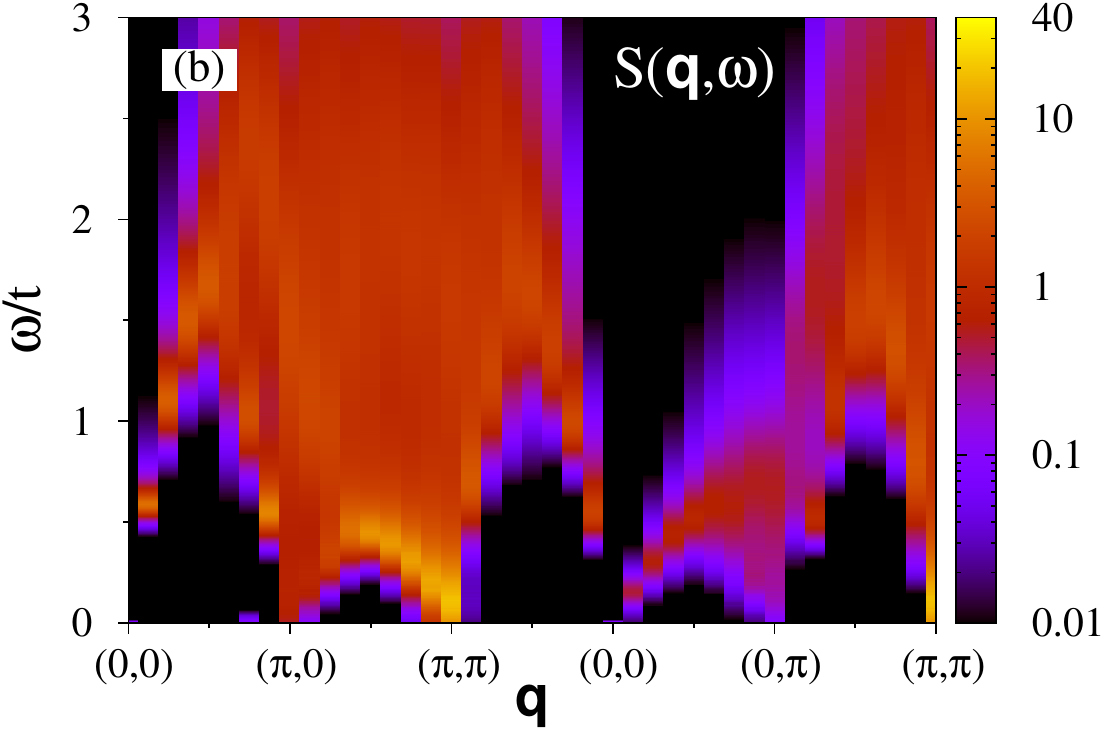}\\
\end{center}
\caption {(Color online) 
Same as in Fig.~\ref{dyn15} but for $t_{\perp}/t=0.3$.
}
\label{dyn3}
\end{figure}

We now examine the spin and charge responses at our largest interchain 
coupling $t_{\perp}/t=0.3$.
A closure of the pseudogap in the single-particle spectral function $A({\pmb k},\omega)$ 
illustrated in Fig.~\ref{G_nod} gives rise to a more pronounced gapless charge mode 
in the long-wavelength limit ${\pmb q}\to 0$, see Fig.~\ref{dyn3}(a).
As follows from Eq.~(\ref{Sigma_Cqw}), this is in turn reflected in a marked increase 
in the Drude-like weight shown in Fig.~\ref{Sig_20} thus providing the evidence of an \emph{increased} 
mobility of charge carriers. Consequently, the low-frequency part of the dynamical spin structure factor 
$S({\pmb q},\omega)$ cannot anymore be interpreted solely in terms of spin-wave-like excitations 
assuming \emph{localized} spins.~\cite{Eder95a,Demler14}  Our next goal is to identify these various   
components in the spin excitation spectrum.

Figure~\ref{dyn3}(b) reveals featureless continua in  $S({\pmb q},\omega)$ near ${\pmb q}=(0,\pi)$ 
and ${\pmb q}=(\pi,0)$ momenta. 
It is natural to assign these incoherent excitations to the continuum of \emph{independent} 
particle-hole pairs, a hallmark of an electron system with mobile charge carriers. 
Indeed,  in the presence of electronic QPs, a \emph{single} spin-flip excitation can be made at 
arbitrary low energy. Hence, the paramagnons cannot propagate without exciting unbound particle-hole pairs. 
The latter contribute to the charge excitation spectrum $C({\pmb q},\omega)$ which features 
a similar continuum around ${\pmb q}=(0,\pi)$, cf. Fig.~\ref{dyn3}(a). This similarity confirms that these features 
stem from the particle-hole bubble of dressed single-particle Green's functions.~\cite{Schmitt12}

Less clear evidence of the particle-hole excitations in $C({\pmb q},\omega)$ 
is found  at ${\pmb q}=(\pi,0)$: since most of the spectral weight is exhausted 
by a high-frequency $\omega/t\simeq 4$ charge mode,~\footnote{Comparison of Figs. \ref{chain} and \ref{dyn3} 
supports the point of view that the high-frequency features of $S({\pmb q},\omega)$ and $C({\pmb q},\omega)$ 
at $t_{\perp}/t=0.3$ are reminiscent of the 1D isolated-chain limit.} 
it is more difficult to resolve the expected particle-hole continuum. 
However, we believe that a faint low-frequency mode corresponds to a lower bound of the particle-hole continuum.

In contrast, the particle-hole interaction vertex remains finite near the AF wave-vector ${\pmb q}=(\pi,\pi)$. 
Consequently, an excited particle and a hole are \emph{bound} together in this part of the Brillouin zone. 
This leads to low-frequency paramagnon excitations, see Fig.~\ref{dyn3}(b).
As compared to the weakly coupled regime shown in Fig.~\ref{dyn15}(b), 
these paramagnons are broadened by scattering off mobile charge carriers 
and dissolve into a FL-like particle-hole continuum on moving away from the AF wave-vector.  
Such a \emph{localized} nature of spin fluctuations in momentum space restricted 
to a narrow range around ${\pmb q}=(\pi,\pi)$ arises from \emph{spatial} AF spin correlations rather than 
from a set of mutually interacting local moments.~\cite{Moriya_book}

\subsubsection{Thermal crossover}

Independently of the intrinsic interest in finding  fingerprints of the LL behavior 
at elevated temperatures, quasi-1D materials often exhibit low-$T$ broken-symmetry ground states. 
These instabilities occur at a temperature scale at which the system effectively becomes 
three-dimensional and long-range order can occur at low but finite temperatures.

The physics associated with a thermal crossover can be studied in weakly coupled 
Hubbard chains: for a weak interchain superexchange coupling $J_{\perp}/J\ll 1$, 
the energy difference between the broken-symmetry AF ground state and excited states 
is small, thus facilitating their thermal population. 
This offers the opportunity to analyze the interaction of low-energy electronic QPs with 
finite-range AF spin fluctuations in a thermally disordered  quasi-1D  metal and to probe transient
changes in the spin excitation spectrum intimately connected to the onset of the
long-range AF order in the $T\to 0$ regime.

\begin{figure}[t!]
\begin{center}
\includegraphics*[width=0.43\textwidth]{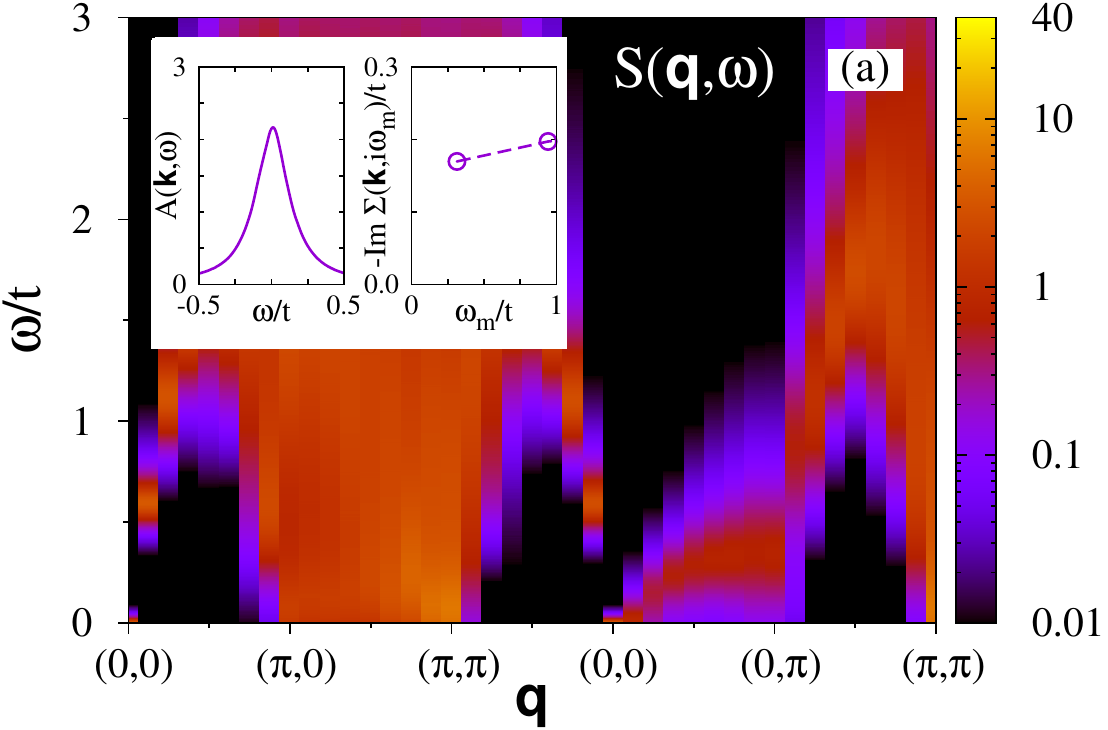}\\
\includegraphics*[width=0.43\textwidth]{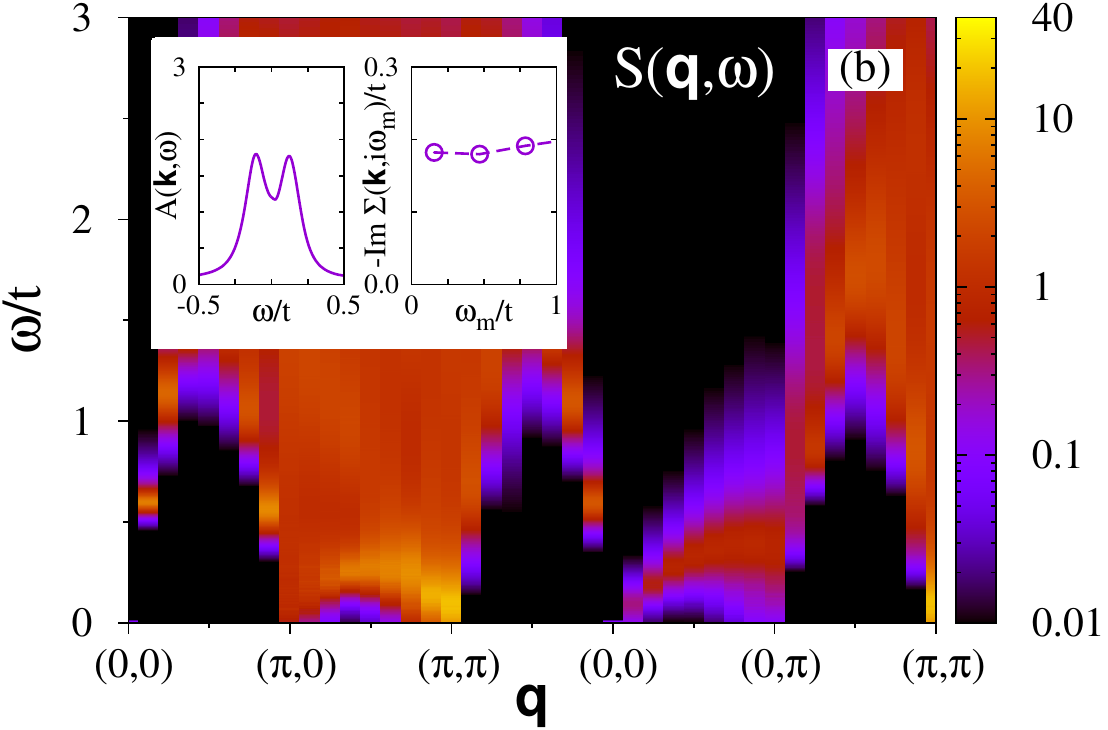}\\
\includegraphics*[width=0.43\textwidth]{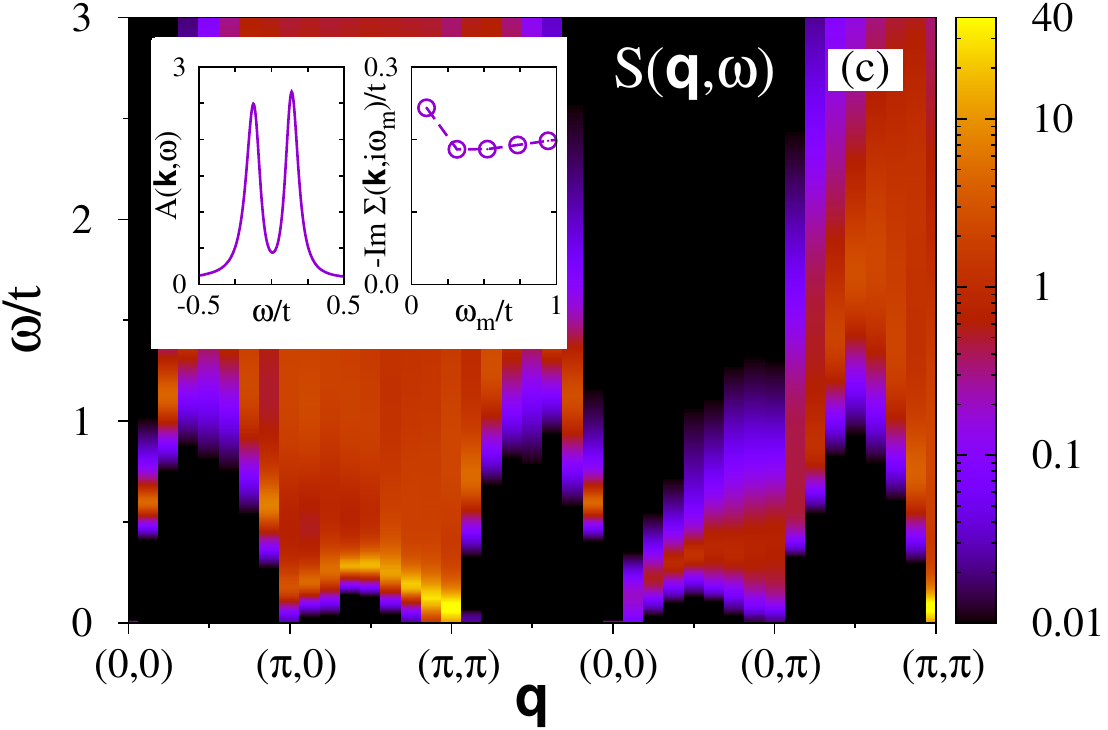}
\end{center}
\caption {(Color online) 
Temperature dependence of the dynamical spin structure factor $S({\pmb q},\omega)$ 
obtained on a $16\times 16$ lattice with  fixed 
$t_{\perp}/t=0.2$: 
(a) $T=t/10$, (b) $T=t/20$, and (c) $T=t/30$.
Insets show the single-particle spectral function at ${\pmb k}=(\pi/2,\pi/2)$
(left) and low-frequency dependence of the imaginary part of the corresponding 
self-energy (right).
}
\label{Sqw_T}
\end{figure}

Figure~\ref{Sqw_T} tracks the temperature dependence of the dynamical spin structure factor 
$S({\pmb q},\omega)$ at a fixed $t_{\perp}/t=0.2$. 
For this value of the interchain coupling, optical spectra in Fig.~\ref{Sig_T} 
provide evidence of a thermal crossover from a FL-like regime at $T=t/10$ 
to the broken-symmetry SDW ground state in the effective zero-temperature limit.

At elevated temperature $T=t/10$, see Fig.~\ref{Sqw_T}(a), $S({\pmb q},\omega)$ features solely 
broad continua associated with single-particle spin-flip excitations -- 
prominent features of the FL dynamics. 
The validity of the FL-like picture is further supported by a well defined QP peak in 
the single-particle spectral function $A({\pmb k},\omega)$ resolved at the nodal ${\pmb k}=(\pi/2,\pi/2)$ point, 
see Fig.~\ref{Sqw_T}(a).  These QP excitations are responsible for a Drude-like feature in 
the corresponding optical conductivity shown in Fig.~\ref{Sig_T}.

Although a Drude-like weight signals that the system remains metallic upon cooling down to $T=t/20$, 
the single-particle dynamics becomes \emph{incoherent}; 
a single QP peak in $A({\pmb k},\omega)$  at the ${\pmb k}=(\pi/2,\pi/2)$ point 
is replaced by two broadened peaks separated by a shallow pseudogap, see Fig.~\ref{Sqw_T}(b). 
This double peak structure yields a \emph{finite}-frequency hump in the 
optical spectra due to particle-hole excitations \emph{across} the pseudogap, cf. Fig.~\ref{Sig_T}.
Further evidence of the growing incoherence in the system is brought by the enhancement 
of the QP scattering rate $\Gamma_{{\pmb k}}=-\textrm{Im}\, \Sigma({\pmb k},i\omega_0)$, 
cf. insets in Figs.~\ref{Sqw_T}(a) and \ref{Sqw_T}(b). 
As is apparent from the low-$T$ growth of the staggered spin structure factor $S(\pi,\pi)$ depicted in Fig.~\ref{S_T}(b), 
the change in the QP damping rate should be traced to the increase in the magnetic correlation length. 
This has a pronounced effect on the spin excitation spectrum which develops a paramagnon branch 
near the AF wave vector, see Fig.~\ref{Sqw_T}(b).  
As indicated by their broad spectral width, these paramagnons have a short lifetime 
due to scattering off mobile charge carriers and merge into a particle-hole continuum on moving 
away from the ${\pmb q}=(\pi,\pi)$ momentum.

\begin{figure}[t!]
\begin{center}
\includegraphics*[width=0.43\textwidth]{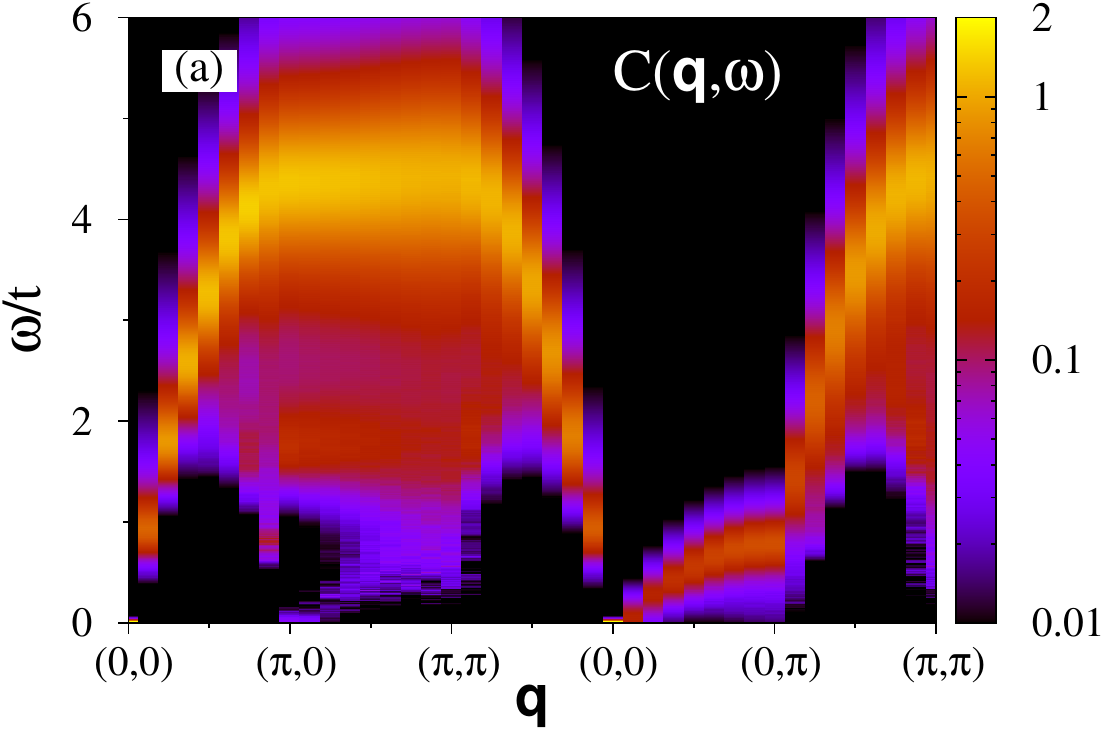}\\
\includegraphics*[width=0.43\textwidth]{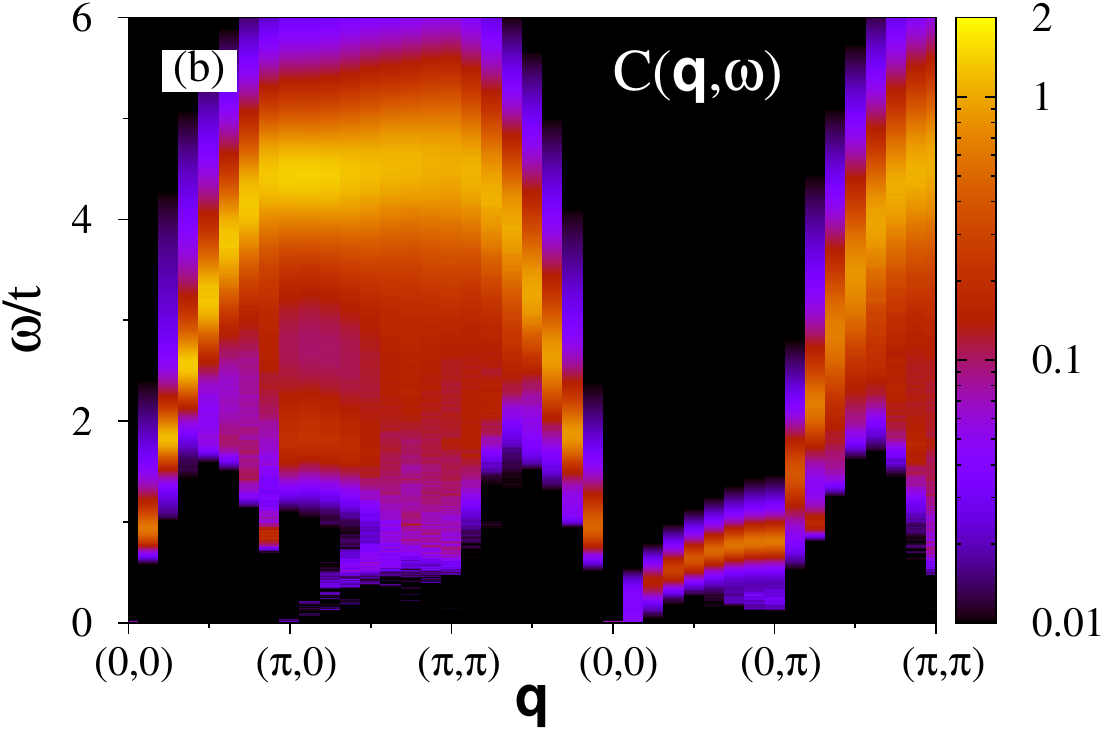}\\
\includegraphics*[width=0.43\textwidth]{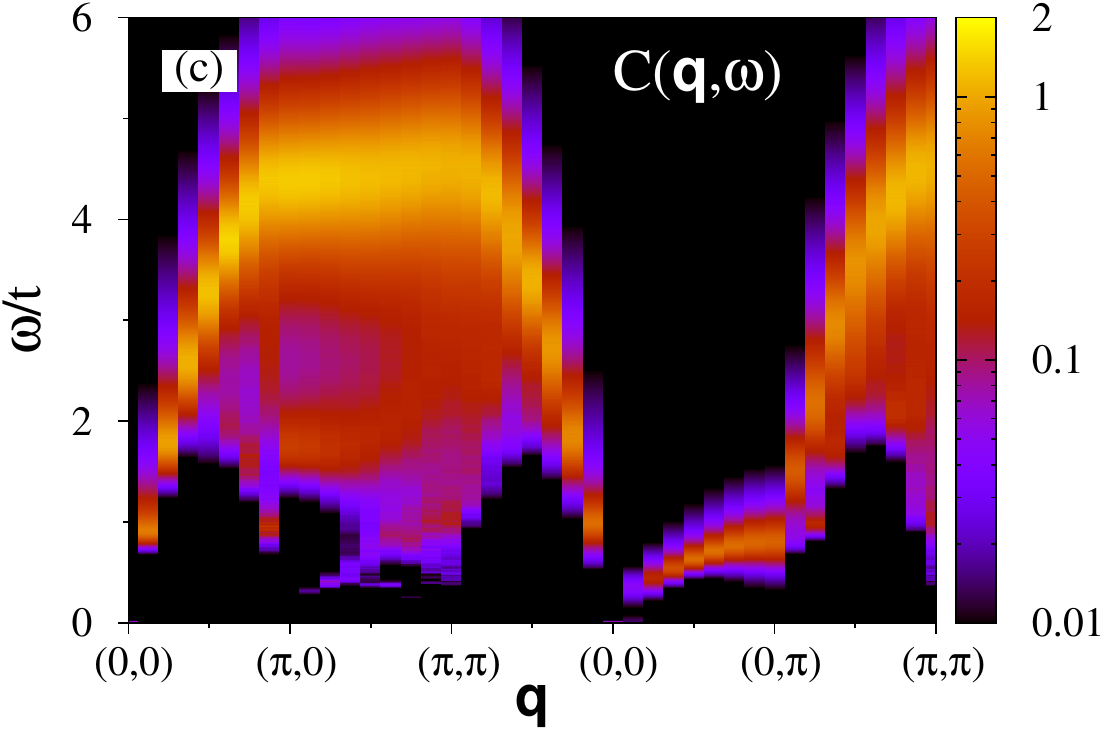}
\end{center}
\caption {(Color online)
Temperature dependence of the dynamical charge structure factor $C({\pmb q},\omega)$
obtained on a $16\times 16$ lattice with  fixed $t_{\perp}/t=0.2$:
(a) $T=t/10$, (b) $T=t/20$, and (c) $T=t/30$.
}
\label{Cqw_T}
\end{figure}

Finally, at our lowest temperature $T=t/30$, a further increase in the QP damping 
rate $\Gamma_{{\pmb k}}$ leads to a strong depletion of single-particle spectral weight, 
see the inset in Fig.~\ref{Sqw_T}(c).
Despite some thermally excited QP states at the Fermi level, a complete suppression 
of the Drude weight in the corresponding optical conductivity shown in Fig.~\ref{Sig_T} 
provides clear evidence of  charge localization.
Hence, the formation of the insulating state appears here as the outcome of a divergent QP mass
rather than the disappearance of charge carriers.

This insulating phase features rich spin dynamics with a pronounced frequency-dependent damping, see Fig.~\ref{Sqw_T}(c). 
On the one hand, we resolve at our lowest temperature $T=t/30$ the increase in both the magnetic intensity 
and stiffness of the spin-wave-like dispersion near the AF wave-vector. 
The latter effect is similar to spin-wave stiffening found in the weakly doped two-dimensional $t$-$J$ model 
on reducing the hole doping and approaching the long-range AF phase.~\cite{Fulde92,Horsch93,Vojta96}  
Moreover, a correlation-induced gap for spin excitations becomes discernible at the 
${\pmb q}=(\pi,0)$ momentum. The size of this gap might be considered as a measure of the vertex correction, i.e,  
binding energy between an excited particle and a hole, a precursor of collective spin excitations.
On the other hand, the interaction of electronic single-particle excitations with 
magnetic modes results in the overdamped spin dynamics at energies above the pseudogap in $A({\pmb k},\omega)$.

Figure~\ref{Cqw_T} summarizes the corresponding  temperature evolution of the dynamical charge structure 
factor $C({\pmb q},\omega)$. The intensity of the low-frequency weight in the long-wavelength 
limit ${\pmb q}\to 0$ progressively decreases with decreasing temperature and is completely washed out 
below the crossover scale marking the onset of the SDW phase. This agrees with a small gap 
in the optical conductivity resolved at $T=t/30$, cf. Fig.~\ref{Sig_T}.

\subsection{Discussion: experimental relevance}
\label{Exp}

Let us now discuss the relevance of our results to quasi-1D organic Bechgaard-Fabre 
salts.~\footnote{The building blocks of this family of compounds are nearly planar organic molecules: 
the selenium-based tetramethylselenofulvalene (TMTSF) or its sulfur analog tetramethyltetrathiafulvalene (TMTTF).}  
At ambient pressure, the Fabre (TMTTF)$_2$X salts are typically Mott insulators 
due to a combined effect of umklapp process and dimerization leading effectively to a half-filled band. 
Upon increasing pressure, the 1D Mott phase is replaced by a higher-dimensional metal when 
the interchain hopping reaches the order of the Mott gap.~\cite{Vescoli98,Pashkin10}  
In contrast, the Bechgaard (TMTSF)$_2$X salts are quasi-1D metals due to larger interchain hopping. 
It effectively warps the FS thus introducing deviation from ideal nesting properties of the 1D limit. 
The optical conductivity spectra of the Bechgaard salts are markedly different from those of a simple metal: 
a narrow Drude-like weight is accompanied by a finite-frequency feature exhausting most of 
the spectral weight.~\cite{Dressel96} 
The latter is usually interpreted in terms of the remnant 1D Mott gap.~\cite{Schwartz98} 
The Drude feature survives up to arbitrarily low temperatures before being 
disrupted by the onset of a small SDW gap driven by the FS nesting.~\cite{Degiorgi96}  

On the one hand, recent CDMFT studies capturing only short-range AF spin correlations have shown 
evidence of a quasi-1D metallic phase with a FS broken into pockets.~\cite{Raczkowski12} 
As the interchain coupling was increased, a continuous FS was restored although with 
a substantial variation of the QP weight in the Brillouin zone. 
A strong momentum dependence of the self-energy offers a simple framework 
accounting for the unusual frequency dependence of the optical conductivity in (TMTSF)$_2$X salts.   
Indeed, it is natural to associate a Drude-like feature (finite-frequency absorption) to the 
FS pockets (gapped regions of the Brillouin zone), respectively. 

On the other hand, our QMC simulations yield a continuous FS in the quasi-1D metallic 
phase. As a result, a two-component optical response follows from the depletion of single-particle spectral 
weight at the Fermi level with a finite-frequency contribution assigned to particle-hole excitations 
across the pseudogap. 
One possible reason behind the continuous FS is that the formation of the FS pockets requires simulations 
in the thermodynamic limit. Alternatively, it is known that the breakup of the FS into 
pockets appears only at sufficiently low temperatures,~\cite{Raczkowski12} i.e., possibly below the onset of the 
intervening SDW phase. In this respect, a larger deviation from the FS nesting, e.g., 
a larger next-nearest-neighbor hopping $-t'$,  would be required to extend the region of stability 
of the quasi-1D metallic phase to lower temperatures.   

Finally, we point out a strong renormalization of the QMC Drude-like weight with respect to the one found 
in the chain-DMFT studies.~\cite{Bier01} It stems from dressing of mobile charge carriers with 
a cloud of finite-range AF spin fluctuations beyond the chain-DMFT approximation. 
Interestingly, a reduced QMC Drude response is reminiscent of the experimental data on the (TMTSF)$_2$X salts 
with the zero-frequency mode containing a tiny $\sim 1\%$ fraction of the total optical weight.~\cite{Dressel96}

\section{Conclusions}
\label{Conclusions}

We have presented a thorough auxiliary-field QMC study of a dimensional crossover in weakly coupled 
1D Hubbard chains with a half-filled band. As a function of the interchain coupling $t_{\perp}$, 
we interpret our finite-temperature results in terms of the crossover from the 1D Mott phase which 
exhibits spin-charge separation to a higher-dimensional AF metal.  
We have placed emphasis on clarifying the nature of this metallic phase and its low-energy excitations. 
To this aim, we have examined both the single- and two-particle excitation spectra. 

On the one hand, the simultaneous emergence of the zero-frequency Drude-like response 
in both intra- $\sigma_{\parallel}(\omega)$ and interchain $\sigma_{\perp}(\omega)$  
optical conductivities implies the onset of a quasi-1D metal at $t_{\perp}/t=0.15$. 
This should be contrasted with the anisotropic localization in coupled spinless LL chains 
leading to metallic transport in some directions and not along the others (smectic metal).~\cite{Emery00} 
On the other hand, most of the optical weight does not form a coherent Drude peak 
as in a FL metal, but accumulates at finite frequency.  This signals unconventional 
charge dynamics attributed to enhanced QP scattering off finite-range AF spin fluctuations. 
Our study reveals a rich structure of the corresponding magnetic excitation spectrum: 
signatures of the FL dynamics (particle-hole continua) and a paramagnon branch near 
the ${\pmb q}=(\pi,\pi)$ momentum  are apparent.

The temperature dependence of the single- and two-particle excitation spectra in the quasi-1D regime 
with $t_{\perp}/t=0.2$ is equally rich.
At elevated temperatures, they are suggestive of the formation of a FL-like metal.
In contrast, at our lowest temperatures, the disappearance of the Drude weight indicates the onset 
of an insulating ground state gapped out by AF spin correlations. 
As a result, we resolved a spin-wave-like mode in the dynamical spin structure factor $S({\pmb q},\omega)$ 
spreading out nearly through the entire Brillouin zone. 

Finally, we have discussed a possible relationship of our findings to the experiments.
On the one hand, the Drude weight in quasi-1D organic Bechgaard-Fabre salts 
arises from deviations from the strictly 1D half-filled band which lead to warping of 
the FS.~\cite{Giamarchi04}  Moreover, there is strong experimental evidence of the importance of higher-dimensional 
AF spin correlations. Thus, the extension of the model Hamiltonian from 1D to quasi-1D is 
essential to account for the metallic behavior in these compounds.
On the other hand, finite-energy absorption in the optical conductivity spectrum indicates 
that the interchain coupling becomes ineffective above a certain frequency threshold. In this regime, 
one recovers the 1D physics with remnant umklapp scattering. 
     
Instead, our QMC data indicates that the high-frequency 1D Mott feature in the optical response 
is quickly replaced upon coupling the  chains by a low-frequency mode associated with the onset of 
finite-range higher-dimensional AF spin correlations. We conclude that the low-temperature optical spectra in quasi-1D 
correlated metals with a half-filled band and a nearly nested FS are intimately related to dominant AF spin fluctuations. 
However, we expect that the importance of longer-range interchain hopping in the actual electronic band 
structure of Bechgaard-Fabre salts~\cite{Jeschke13,Sakai13} will enhance the frustration effects and thus could 
have the potential of reproducing the experimentally observed optical spectra in lattice simulations.

\begin{acknowledgments}
We thank the LRZ-Munich and the J\"ulich Supercomputing Centre for a generous 
allocation of CPU time.   
This work was supported by the FP7/ERC Starting Grant No. 306897 and FP7/Marie-Curie Grant No. 321918. 
F.~F.~A. acknowledges support from the DFG grant AS120/8-2 (FOR1346).
\end{acknowledgments}

\bibliographystyle{apsrev4-1}

%

\end{document}